\documentclass[aps,floatfix,prd,letterpaper,twocolumn,nofootinbib]{revtex4} 
\usepackage{graphicx,amsmath,amssymb,color,bm}

\usepackage{hyperref}
\def\be{\begin{equation}}
\def\ee{\end{equation}}
\def\bea{\begin{eqnarray}}
\def\eea{\end{eqnarray}}
\newcommand{\vs}{\nonumber\\}
\def\ba#1\ea{\begin{align}#1\end{align}}
\def\bg#1\eg{\begin{gather}#1\end{gather}}

\def\Mpch{\,h^{-1}\,{\rm Mpc}}
\def\iMpch{\,h\,{\rm Mpc}^{-1}}

\newcommand{\s}{\sigma}
\newcommand{\refeq}[1]{Eq.~(\ref{eq:#1})}          
\newcommand{\refeqs}[2]{Eqs.~(\ref{eq:#1})--(\ref{eq:#2})}          
\newcommand{\reffig}[1]{Fig.~\ref{fig:#1}}          

\newcommand{\refapp}[1]{App.~\ref{app:#1}}
\newcommand{\refsec}[1]{Sec.~\ref{sec:#1}}

\renewcommand{\v}[1]{\bm{#1}}

\newcommand{\vx}{\v{x}}
\newcommand{\vy}{\v{y}}
\newcommand{\vv}{\v{v}}
\newcommand{\vk}{\v{k}}
\newcommand{\vp}{\v{p}}
\newcommand{\vq}{\v{q}}

\newcommand{\<}{\langle}
\renewcommand{\>}{\rangle}

\renewcommand{\d}{\delta}

\newcommand{\eps}{\epsilon}

\newcommand{\Om}{\Omega_m}

\newcommand{\rhob}{\bar\rho}
\def\G{\mathcal{G}}
\def\O{\mathcal{O}}

\def\cH{\mathcal{H}}

\begin{document}

\title{The effect of relative velocity and density perturbations between baryons and dark matter on the clustering of galaxies}

\author{Fabian Schmidt}
\affiliation{Max-Planck-Institut f\"ur Astrophysik, Karl-Schwarzschild-Str.~1, 85748~Garching, Germany}

\begin{abstract}
Pre-recombination acoustic oscillations induce non-adiabatic perturbations between baryons and dark matter, corresponding to a constant relative-density $\d_{bc}$ and decaying relative-velocity perturbation $\vv_{bc}$.  Due to their significant large-scale correlations and prominent baryon acoustic oscillation (BAO) features, these modes are potentially important for the use of the BAO as standard ruler.  We present a complete treatment of the effects of the baryon-CDM perturbations on galaxy clustering in the context of a rigorous perturbative bias expansion.  The leading effects are proportional to $\d_{bc}$ and $\theta_{bc} = \partial_i v_{bc}^i$.  We estimate the magnitude of these terms through the excursion set approach.  The contribution from $v_{bc}^2$, which has attracted significant attention recently, contributes at subleading (1-loop) order.  
The relative-density contribution $\d_{bc}$ is expected to be by far the largest contribution.  
We also point out contributions to the galaxy velocity bias, the largest of which is simply $v_{bc}$, leading to a term $\propto \mu^2\theta_{bc}$ in the redshift-space galaxy power spectrum $P_g^s(k,\mu)$.  Complete expressions of the galaxy power spectrum at 1-loop order are given, which contain several new terms.  
\end{abstract}

\date{\today}

\maketitle

\section{Introduction}
\label{sec:intro}

Our Universe contains two dominant matter components: cold dark matter (CDM, $c$)
and baryons ($b$, i.e. all non-relativistic standard model particles).  
In studies of structure formation, we commonly treat these two fluids as
a single, comoving matter fluid (or, a collection of collisionless particles).  However, the coupling of baryons to radiation in the primordial plasma before recombination leads to relative perturbations in density and velocity of the baryon and CDM components.  While significant initially, these perturbations grow less rapidly than the adiabatic growing mode and hence are very small in the low-redshift universe.  However, since these perturbations have significant large-scale correlations, they are potentially detectable through their imprint in the clustering of galaxies on large scales.  Moreover, they retain a significantly stronger imprint of the BAO feature than the adiabatic growing mode, so that they are of relevance for the use of the BAO feature in the galaxy two-point function as a standard ruler \cite{BAO/SDSS,BAO/2dF}.  

Let us consider the evolution of baryons and CDM after baryon-photon decoupling, 
approximating both as pressureless fluids, which is appropriate
on sufficiently large scales.  The evolution of two such fluids coupled by gravity is described by the Euler and continuity equations.  Restricting to
linear order in perturbations (the fully nonlinear case will be discussed in \refsec{nonlin}), these simply become
\ba
\frac{\partial}{\partial\tau}\d_s =\:& -\theta_s \,,\quad s \in \{b,\,c\} \vs
\frac{\partial}{\partial\tau} \theta_s + \cH \theta_s =\:& - \frac32 \Om(a) \cH^2 \d_m\,,
\label{eq:deltalinearEOM1}
\ea
where $\d_s \equiv \d\rho_s/\bar\rho_s$ is the fractional density perturbation in species $s$, $\theta_s = \partial_j v_s^j$ is the fluid velocity divergence of species $s$, while $\d_m = (\Omega_b \d_b + \Omega_c \d_c)/\Omega_m$ is the total matter density perturbation.  Further, $\tau$ denotes conformal time defined through $dt = a d\tau$, and $\cH = d\ln a/d\tau= a H$ is the conformal Hubble rate.  Throughout we work in units where the speed of light $c=1$.  
We also use the density parameters today $\Omega_s = 8 \pi G \bar\rho_s(t_0)/(3 H_0^2)$, and the total matter density parameter $\Omega_m = \Omega_b+\Omega_c$.  Finally, the baryon fraction is $f_b = \Omega_b/\Om$.  It is useful to combine these equations and to rewrite them in terms of $\d_m$ and the \emph{relative} fractional density perturbation $\d_r = \d_b-\d_c$:
\ba
\frac{\partial^2}{\partial\tau^2} \d_m + \cH \frac{\partial}{\partial\tau}\d_m - \frac32 \Om(a) \cH^2 \d_m
=\:& 0 \vs
\frac{\partial^2}{\partial\tau^2} \d_r + \cH \frac{\partial}{\partial\tau}\d_r =\:& 0\,.
\label{eq:deltalinearEOM2}
\ea
Clearly, at linear order the evolution of the total matter density
perturbation $\d_m$ and the relative density perturbation between the
two fluids decouples.  This is not surprising, as the only relevant force
included here is gravity, which does not care about $\d_r$ by way of the
weak equivalence principle.  
We now immediately obtain the general solution of these two decoupled ODE as 
\ba
\d_m(\tau) =\:& A_+ D_+(\tau) + A_- H(\tau) \vs
\d_r(\tau) =\:& R_+ + R_- D_r(\tau)\,,
\label{eq:drgen}
\ea
where $A_\pm,\, R_\pm$ are constants and
\be
D_r(\tau) \equiv H_0^{-1} \int_\tau^\infty \frac{d\tau'}{a(\tau')} 
= \int_{\ln a(\tau)}^\infty \frac{d\ln a'}{a^{\prime 2} H(a')/H_0} 
\,.
\ee
Note that during matter domination, where the universe is closely approximated by the Einstein-de Sitter solution, this approaches $D_r = -2 a^{-1/2}$.  

We see from \refeq{drgen} that $\d_m$ contains the two well-known growing and decaying modes $\propto A_\pm$ of adiabatic perturbations \cite{bernardeau/etal:2002}.  A third mode $\propto R_+ \equiv \d_{bc}$
is a constant compensated perturbation $\d\rho_c = -\d\rho_b$, corresponding to $\d_m=0$ while $\d_r \neq 0$ \cite{barkana/loeb:11,grin/dore/kamionkowski,soumagnac/etal:16}.  This mode can be seen as modulating the local baryon-CDM ratio,
$(\Omega_b/\Omega_c)_{\rm loc} = \overline{\Omega_b/\Omega_c} (1 + \d_{bc})$.  
The significance of the fourth decaying mode $\propto R_-$
becomes clear by considering
\ba
\frac{\theta_{bc}}{\cH} \equiv \frac{\theta_b-\theta_c}{\cH} = R_- \frac{H_0}{a^2 H(a)}\,,
\quad
R_- = \frac{\theta_{bc,0}}{H_0}\,,
\label{eq:Rminus}
\ea
where $\theta_{bc,0} \equiv \theta_{bc}(z=0)$.  
Thus, this mode corresponds to an initial relative velocity $\v{v}_{bc} \propto 1/a$ between the two fluids.  

With few exceptions \cite{shoji/komatsu,somogyi/smith:2010,bernardeau/vdr/vernizzi,lewandowski/perko/senatore}, studies of structure formation using perturbation theory and N-body simulations have focused on the adiabatic growing mode $A_+$.  Ref.~\cite{tseliakhovich/hirata:2010} pointed out that pre-recombination plasma waves (baryon acoustic oscillations, BAO) lead to a significant relative velocity $\v{v}_{bc}$ (also referred to as streaming velocity) at the epoch of baryon-photon decoupling $\tau_{\rm dec}$.  This can leave an imprint in low-redshift structure which assembled out of low-mass halos at high redshifts \cite{dalal/etal:2010,yoo/dalal/seljak,yoo/seljak,slepian/eisenstein,blazek/etal:15}.  
Similarly, the $R_+$-mode is also sourced during recombination \cite{barkana/loeb:11}.  Both $R_-$ and $R_+$ have significant large-scale correlations and in particular large BAO features (\reffig{Pklin}).   For convenience, we will refer to the constant relative density mode $R_+$ and the decaying relative velocity mode $R_-$ jointly as \emph{baryon-CDM perturbations} in the following.  

Since galaxy formation depends sensitively on both baryons and CDM, it is crucial to include these modes when making predictions for galaxy clustering.\footnote{Throughout this paper, we ignore the decaying mode $A_-$, since it is not sourced significantly by recombination physics and decays very rapidly.}  The goal of this paper is to provide a recipe for a complete description of these effects within a rigorous perturbation theory approach (renormalized bias expansion \cite{mcdonald:2006,assassi/etal,MSZ}, which can be seen as an effective field theory \cite{baumann,senatore:2014}).  As an example, we derive the galaxy auto and cross power spectra including the leading nonlinear (1-loop) correction.  Moreover, we provide quantitative estimates of the bias parameters that control the magnitude of the baryon-CDM effects on galaxy clustering.

Previously, of the baryon-CDM perturbations introduced above, only a perfectly uniform streaming
velocity $\v{v}_{bc}$ has been considered in detail, that is $\theta_{bc}$ was set to zero (but see \cite{ahn}).  
$\v{v}_{bc}$ itself can only enter at second order in the galaxy density, since it is a vector, so that the leading contribution to the fractional galaxy number density perturbation is
\cite{dalal/etal:2010,yoo/dalal/seljak,yoo/seljak,slepian/eisenstein,blazek/etal:15}
\be
\d_g(\vx,\tau) \equiv \frac{n_g(\vx,\tau)}{\bar n_g(\tau)} -1
\supset b^{bc}_{v^2}(\tau) \left[\v{v}_{bc}^2 - \s_{v_{bc}}^2(\tau) \right]\,,
\label{eq:dgv2}
\ee
where $\s_{v_{bc}}^2(\tau) \equiv \<\v{v}_{bc}^2\>$.  In order to assess the quantitative
impact of the term in \refeq{dgv2}, we need an estimate for the bias
parameter $b^{bc}_{v^2}$.  Ref.~\cite{dalal/etal:2010} argued that $\v{v}_{bc}$
increases the effective sound speed $c_s$ of the neutral gas, so that the Jeans mass $M_J$ increases by a factor $[1 + v_{bc}^2/c_s^2]^{3/2}$.  
This leads to large effects on low-mass halos prior to reionization, as investigated using small-box simulations in \cite{tseliakhovich/barkana/hirata,Visbal/etal:12,popa/etal}.    
However, how these are transferred to the number density of galaxies at low redshifts 
is unclear.  $b^{bc}_{v^2}$ could be as small as $\sim 10^{-5} \s_{v_{bc}}^{-2}$ \cite{blazek/etal:15,tseliakhovich/barkana/hirata}.  The fiducial value adopted in previous studies is \cite{dalal/etal:2010,yoo/dalal/seljak,yoo/seljak,blazek/etal:15},
\be
b^{bc}_{v^2} \sim 0.01\, \s_{v_{bc}}^{-2}(z) \approx 9.2\times 10^{11} \,(1+z)^{-2}\,.
\label{eq:bv2}
\ee
at redshifts $z \lesssim 2$. Note that given the non-detection in current data, $b^{bc}_{v^2}$ cannot be much larger than this \cite{yoo/seljak}.  
While \refeq{dgv2} can only contribute to the galaxy power spectrum at 1-loop order, the quantities $\d_{bc},\,\theta_{bc}$ discussed above, corresponding to the modes $R_+,\,R_-$, respectively, enter at linear order, and are thus the leading effects unless their bias parameters are highly suppressed compared
to $b_{v^2}$.

\begin{table}[t]
\centering
\begin{tabular}{l|l}
\hline
\hline
$\d_s$ & Density perturbation of species $s$ \\
$\d_m = f_b \d_b + (1-f_b) \d_c$ & Total matter density perturbation \\
$\d_r = \d_b - \d_c$~~[\refeq{drgen}]  & Relative density perturbation \\
$\vv_r = \vv_b-\vv_c$ & Relative velocity \\
\hline
$R_+ \equiv \d_{bc}$ & Initial amplitude of constant \\
& relative density perturbation \\
$R_- \equiv H_0^{-1} \partial_iv_{bc,0}^i$ & Initial amplitude of decaying \\
 & relative velocity perturbation \\
\hline
\end{tabular}
\caption{Notation for baryon-CDM perturbations adopted in the paper.  Actual relative density and velocity perturbations are denoted with a subscript $r$, while their initial amplitudes (extrapolated to $z=0$ using linear theory) are denoted with a subscript $bc$.}
\label{tab:notation}
\end{table}

So far, we have neglected radiation and anisotropic stress, which is sufficiently accurate at $z \lesssim 20$ but not at higher redshifts.   While not important for the formation of structure, this approximation becomes important if one uses transfer function outputs from Boltzmann codes at higher redshifts.  Instead, one should match the modes $A_+,\,R_\pm$ from the transfer function output at low redshift.  We discuss this in \refapp{Tk}.  

The outline of the paper is as follows.  We begin with the leading, linear galaxy power spectrum on large scales in \refsec{biaslin}, and present several estimates for the bias parameters $b_\theta^{bc},\,b_\d^{bc}$ appearing at this order.  In \refsec{nonlin}, we show how a general bias expansion can be constructed up to any desired order.  We also give the complete expressions for the galaxy-auto and cross power spectra at 1-loop order, performing a renormalization of the bias parameters in the process.  \refsec{vel} discusses the baryon-CDM contributions to the galaxy velocity field, which is important for redshift-space distortions.  We conclude in \refsec{concl}.  In the appendix, we discuss how to obtain the transfer functions for $\d_{bc},\,\vv_{bc}$, and provide some details on the spherical collapse calculation used.

Our notation is summarized in Table~\ref{tab:notation}.  
Throughout we set $c=1$, and, for numerical results, assume a flat
$\Lambda$CDM cosmology with $\Omega_m=0.27$, $h=0.7$, $\Omega_b h^2=0.023$, $n_s=0.95$, $\sigma_8=0.791$, and transfer functions given by CAMB \cite{camb} (see \refapp{Tk}).  This yields for the RMS streaming velocity
$\s_{v_{bc}}(z) \equiv \<\v{v}_{bc}^2\>^{1/2} = 0.031 (1+z)\,{\rm km/s}$.

\section{Galaxy clustering and baryon-CDM perturbations: linear order}
\label{sec:biaslin}

Our goal is to write the galaxy density perturbation as
\be
\d_g(\vx,\tau) = \sum_O b_O(\tau) O(\vx,\tau)\,,
\label{eq:dggen}
\ee
where $b_O(\tau)$ are bias parameters while $O(\vx,\tau)$ denote operators (statistical fields).  
Since the galaxy density is a (3-)scalar, the operators $O$ also have to be scalar.  Moreover, we classify terms in the expansion in \refeq{dggen} by the
order in perturbation theory of each operator.   Then, assuming Gaussian
initial conditions, only linear order operators need to be included to
obtain the leading expression for the galaxy two-point function, which we will
present in \refsec{lin}.  
On the other hand, the leading nonlinear correction to the galaxy two-point
function in perturbation theory requires all terms up to third order.  
We stress that, while we only provide rough estimates for the expected magnitude
of the bias parameters $b_O$ here, the list of terms appearing in \refeq{dggen}
is dictated by the two conditions of symmetry and being locally observable and
is thus unambiguous.  That is, if all bias parameters are allowed to be free,
the predictions from \refeq{dggen} are guaranteed to accurately describe 
the data on scales where perturbation theory is valid.  

It is clear from our discussion in \refsec{intro} that, at linear order, we have to allow
for $\d_g$ to depend on all modes of the baryon-CDM fluid system.  This results in three terms:
\ba
\d_g^{(1)}(\vx,\tau) =\:& b_1(\tau) \d_m^{(1)}(\vx, \tau) + b_{R+}(\tau) R_+(\vq[\vx,\tau],\tau) \vs
& + b_{R-}(\tau) R_-(\vq[\vx,\tau],\tau) \label{eq:dg2}\\
=\:& b_1(\tau) \d_m^{(1)}(\vx, \tau) + b_\theta^{bc}(\tau) \theta_{bc}(\vq, \tau) + b_\d^{bc}(\tau) \d_{bc}(\vq)\,,
\nonumber
\ea
where in the second line we have used \refeq{Rminus} and defined $\d_{bc} \equiv R_+$ [whereas $\d_r$ receives contributions from both $\d_{bc}$ and $\theta_{bc}$, \refeq{drgen}].  Note that $R_\pm,\,\theta_{bc},\,\d_{bc}$ [as well as $\vv_{bc}$ in \refeq{dgv2}] are to be evaluated at the Lagrangian position $\vq[\vx,\tau]$ corresponding to $(\vx,\tau)$
\cite{blazek/etal:15}.\footnote{The precisely correct argument is the position of the fluid at $\tau_{\rm dec}$.  However, this distinction makes a negligible difference, of the same order as other nonlinear terms at recombination not considered here.}  At linear order we can neglect this distinction, but it will reappear in \refsec{nonlin}.  Again, a quadratic term such as \refeq{dgv2} enters the galaxy two-point function only at subleading (1-loop) order.  Refs.~\cite{barkana/loeb:11,soumagnac/etal:16} considered the term $\d_{bc}$, while \cite{blazek/etal:15} introduced the term $\theta_{bc}$, albeit with a coefficient which we will argue to be unphysical in \refsec{bthetabv2}.  

In order to assess the quantitative importance of the new terms in
\refeq{dg2}, we need estimates for the bias parameters $b_\theta^{bc},\,b_\delta^{bc}$.  In the next sections, we will discuss several such estimates.  

\subsection{Bias estimate 1: local Eulerian biasing}
\label{sec:thresh}

We begin by considering a simple toy model, namely local Eulerian biasing \cite{fry/gaztanaga:1983,kaiser:1984}.  
Let us assume the galaxy density is a simple local function of the CDM and
baryon densities $\rho_c(\vx, z)$, $\rho_m(\vx, z)$, smoothed on some small scale that is not relevant for large-scale statistics.  We can thus write
$n_g(\vx, z) = F_g[\d_m(\vx, z), \d_r(\vx, z)]$.  
The linear bias w.r.t $\d_m$ is defined as the response of $n_g$ to a long-wavelength perturbation in the total matter, and in this model is thus  given by 
\be
b_1 = \frac1{F_g[0]} \frac{\partial F_g}{\partial\d_m}\Big|_0\,.
\ee
Similarly, we define
\be
b_r = \frac1{F_g[0]} \frac{\partial F_g}{\partial\d_r}\Big|_0\,,
\ee
which essentially quantifies the response of the galaxy density to 
a change in the local baryon-CDM ratio.   We then trivially have
$b_\d^{bc} = b_r$.  Since a change in the local baryon-CDM ratio 
changes the total baryonic mass available to form stars, we expect
$b_r$ to be of order one for real galaxies.  In the remainder of the paper,
we will choose $b_\d^{bc}=1$ as fiducial value.

In order
to derive $b_\theta^{bc}$, we note that a nonzero $\theta_{bc}(\vx,z)$ 
is associated with a nonzero relative density perturbation derived above, given by \refeq{drgen},
\ba
\d_r\Big|_{\theta_{bc}} = \frac{\theta_{bc,0}}{H_0} D_r(\tau)
\approx -2 \frac{\theta_{bc}(z)}{H_0} (1+z)^{-1/2}\,,
\ea
where the second relations holds in matter domination.  We then have
\be
b_\theta^{bc} = \frac1{\bar n_g} \frac{\partial\bar n_g}{\partial\theta_{bc}(z)}\Big|_0
= b_r \frac{\partial \d_r}{\partial\theta_{bc}(z)} = -2 (1+z)^{-1/2} H_0^{-1}\,b_r\,.
\ee
Thus, assuming $b_r$ is of order one, $b_\theta^{bc}$ is estimated to be of order $H_0^{-1}$.

\subsection{Bias estimate 2: excursion set}
\label{sec:scollapse}

In order to improve upon this model, we consider the excursion set 
approach \cite{bond/etal:91}:  the abundance of halos at fixed mass
is proportional to the probability of the initial density field at a given point first crossing a density threshold
$\d_{\rm crit}$ when lowering the smoothing scale from infinity down to the Lagrangian 
radius $R(M)$ of those halos.  The mean number density of halos $\bar n_h(M)$ is then
a function of $\nu_c = \d_{\rm crit}(z)/\sigma(M,z)$, where $\sigma(M,z)$
is the square root of the variance of the linear matter density field
at redshift $z$ smoothed on the scale $R(M)$.  The collapse threshold
$\d_{\rm crit}(z) \approx 1.7$ is usually derived as the linearly extrapolated initial 
spherical overdensity that, when followed fully nonlinearly, collapses 
at redshift $z$.  A long-wavelength matter density perturbation $\d_\ell$ (in the
growing mode) is locally equivalent to reducing the threshold $\d_{\rm crit} \to \d_{\rm crit} - \d_\ell$, from which we obtain a prediction for the Lagrangian bias \cite{cole/kaiser:1989,mo/white:1996}:
\be
b_1^L = b_1 - 1 = -\frac1{\bar n_h(M)} \frac{\partial\bar n_h(M)}{\partial\d_{\rm crit}}\,.
\ee
Thus, if we can derive how the collapse threshold changes under a long-wavelength perturbation in $\theta_{bc}$ and $\d_{bc}$, we obtain an estimate for $b_\theta^{bc}$ via
\be
b_\theta^{bc} = \frac1{\bar n_h} \frac{\partial\bar n_h}{\partial\theta_{bc}}
= - b_1^L \frac{\partial\d_{\rm crit}}{\partial \theta_{bc}}\,,
\label{eq:bthetab1}
\ee
and analogously for $\d_{bc}$.

In order to derive $\d_{\rm crit}$ in the presence of baryon-CDM perturbations, we consider the following
setup.  We follow a spherical perturbation in the CDM component $\d_c$, which eventually collapses to form a halo at late times.  Assuming that Silk damping has erased density perturbations in the baryon component, we take it to be uniform, $\d_b=0$, until decoupling at $z(\tau_{\rm dec})\simeq 1000$; this is a good approximation for the small-scale perturbations that collapse to halos.  Note that this does not erase the effect of \emph{large-scale} perturbations in the baryon-CDM ratio $\d_{bc}$, which are still present in the large-scale environment.  
After decoupling, the baryon fluid is pressureless, but has a \emph{uniform velocity divergence} relative to dark matter;  here we set $\v{v}_{bc}=0$, since at leading order the effects of $\v{v}_{bc}^2$ and $\theta_{bc}$ decouple.  
Ref.~\cite{asaba/ichiki/tashiro} study the evolution of a tophat perturbation in the presence of a uniform relative velocity $\v{v}_{bc}$.

\begin{figure}[t!]
\centering
\includegraphics[trim=1.8cm 7.6cm 2.8cm 1.5cm,clip,width=0.49\textwidth]{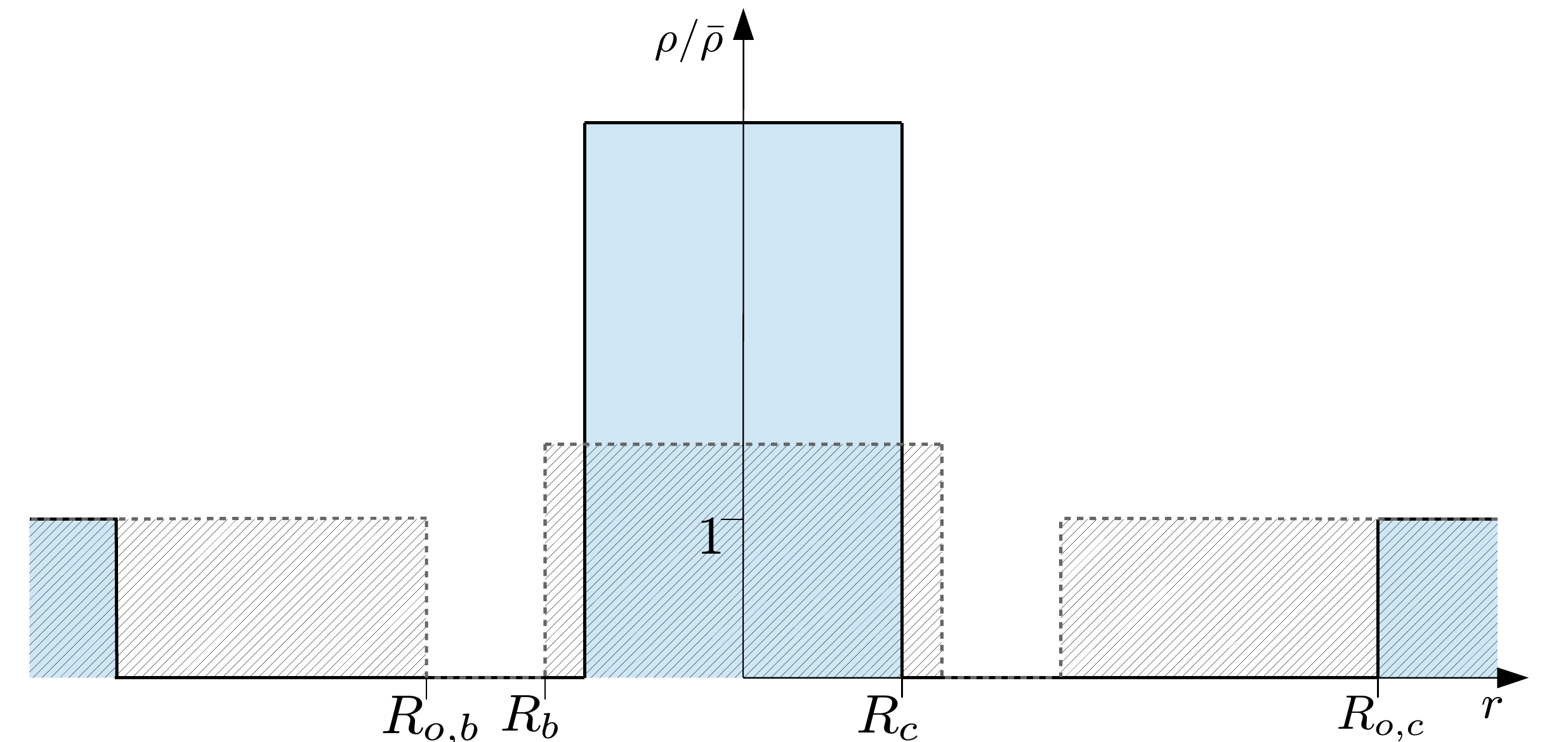}
\caption{Illustration of the two-fluid (CDM, solid; baryons, dashed/hatched) spherical collapse setup considered here.  The tophat shell radii $R_b,\,R_c$ and compensation radii $R_{o,b},\,R_{o,c}$ are indicated.  The shells have the same initial radii at $\tau_{\rm dec}$, but different overdensities.  In the presence of a relative velocity divergence, $R_b$ and $R_c$ evolve differently.  
\label{fig:sketch}}
\end{figure}

We further restrict to a tophat (uniform density) perturbation $\d_s(\tau)$
of radius $R_s(\tau)$, $s=c,b$.  Each shell is assumed to be surrounded by space devoid of the corresponding matter component (see \reffig{sketch}) out to a compensation radius $R_o$, where
\be
R_{o,s}(\tau) = [1 + \d_s(\tau)]^{1/3} R_s(\tau)\,.
\label{eq:Ro}
\ee
Then, the evolution equations reduce to a generalization of the familiar spherical collapse equation,
\ba
& \frac{\ddot R_s}{R_s} = -\frac{4\pi G}3 \left[\rhob - 2 \rho_\Lambda \right] - G_s\,,
\label{eq:Rbc}\\
& G_s = \sum_{t=c,b} f_t 
\left\{
\begin{array}{cc}
\d_t \,, & R_t \leq R_s \\[3pt]
{\rm max}\{0, (1+\d_t) (R_s/R_t)^3 - 1 \}\,, & R_t > R_s \,,\\
\end{array} \right. 
\nonumber
\ea
where dots denote derivatives with respect to time $t$ and $f_t = \Omega_t/\Om$.  
Further, during the tophat evolution mass is conserved so that
\be
1+\d_s(\tau) = [1+\d_s(\tau_{\rm in})] 
\left( \frac{[R_s/a](\tau)}{[R_s/a](\tau_{\rm in})}\right)^{-3}
\,.
\label{eq:delta_i}
\ee
Note that the evolution of two tophat shells with different radii is not consistent; that is, the density profiles will not remain tophat due to the mutual gravitational interaction.  We will ignore this effect and simply follow the two shells here. Our results are thus approximate, but sufficient for order-of-magnitude forecasts.  

We integrate the equations for $R_b,\,R_c$ starting from $\tau_{\rm in}=\tau_{\rm dec}$.  This calculation and the initial conditions, which involve some subtleties, are described in \refapp{scollapse}.   The initial conditions need some care, since an initial relative velocity divergence $\theta_{bc}$ can induce an unphysical $R_+$ mode if the latter is not properly nulled.  Further, while the overall scale of the tophat perturbation is arbitrary, the ratio $R_b(\tau_{\rm in})/R_c(\tau_{\rm in})$ is not.  After decoupling, all baryons within the CDM compensation radius $R_{o,c}$ [\refeq{Ro}] depart from their initial velocities and begin to collapse onto the CDM perturbation.  Thus, it would seem sensible to choose $R_b(\tau_{\rm in}) = R_{o,c}(\tau_{\rm in})$.  However, one can easily verify that we do not recover linear evolution from the initial conditions unless $R_b(\tau_{\rm in}) = R_c(\tau_{\rm in})$, an artefact of fixing the tophat shape of the shells.  Since the collapse threshold is only meaningful if the tophat density initially follows linear evolution, we thus need to choose $R_b(\tau_{\rm in}) = R_c(\tau_{\rm in})$ (as chosen in \cite{asaba/ichiki/tashiro,ichiki/takada}).  
We then adjust $\d_{c,\rm in}$ so that the CDM shell reaches
$R_c=0$ at the desired redshift $z_{\rm coll}$; following the baryon shell leads to identical results, since the two are comoving in the late stages of collapse.  The collapse threshold $\d_{\rm crit}$ is obtained by
integrating the linear growth equation for $\d_m$ with the same initial
conditions to the same redshift $z_{\rm coll}$.  Note that one has to use 
the linear total matter perturbation since this is what $\sigma(M,z)$ refers to in the definition of $\nu_c = \d_{\rm crit}/\sigma(M,z)$.

We begin with the case $\d_{bc}\neq0$ while $\theta_{bc}=0$;  this can be implemented by changing $f_b$ in \refeq{Rbc}.  By construction, $R_b=R_c$ initially while in this case we also have $\dot R_b = \dot R_c$.  From \refeq{Rbc} we see that the force acting on both shells is the same and hence $R_b(\tau)=R_c(\tau)$ holds at all times.  This means that the $R_+$-mode $\d_{bc}$ remains constant not only at linear order, but throughout the nonlinear tophat evolution.  Hence, a constant compensated density perturbation does not change the 
collapse threshold,  $\partial\d_{\rm crit}/\partial\d_{bc}=0$.  This is
again an artefact of the double tophat;  one can show that $\d_r$ no longer
remains constant at second order in perturbation theory [i.e., by solving
\refeq{deltanlEOM} below].  In any case,
the local baryon-CDM ratio is still modified, and we expect a order unity
$b_\d^{bc}$ for galaxies following \refsec{thresh}.

We now turn to the response of $\d_{\rm crit}$ to $\theta_{bc}$.  In this case,
even though we start with $R_b=R_c$ at $\tau_{\rm dec}$, we have $\dot R_b - \dot R_c \propto \theta_{bc}$ so that the shells evolve differently.  This in turn leads to a nontrivial gravitational coupling between the two shells which influences the collapse.  \reffig{dcrit} in \refapp{scollapse} (filled triangles) shows $\d_{\rm crit}$ as a function of $\theta_{bc}$.  We fit a linear relation in the range 
$|\theta_{bc,0}/H_0| \leq 2\cdot 10^{-7}$, and obtain
\be
\frac{\partial\delta_{\rm crit}}{\partial(\theta_{bc,0}/H_0)} = -6.8 \quad\mbox{at}\  z=1.2\,,
\label{eq:dcrit}
\ee
with a very small redshift evolution from $-7.1$ at $z=0$ to $-6.4$ at $z=10$.\footnote{This result is smaller by a factor $\sim 10$ than the number given in a previous arXiv version of this paper. The differences resulting in this change are described in \refapp{scollapse}.}  
The slope is negative, meaning that the collapse threshold is lower when baryons fall onto the overdensity with a smaller initial infall velocity than the CDM which is in the growing mode.  This might seem counterintuitive.  Note however that the growth suppression due to $\theta_{bc}$ is also contained in the linear growth used to extrapolate $\d_{\rm crit}$ to low redshifts.  Thus, the negative slope says that the fractional suppression in the nonlinear growth is smaller than that in the linear growth (similar results have been found for modified gravity in e.g. \cite{schmidt/hu/lima}, where growth is enhanced but $\d_{\rm crit}$ is reduced).  

\refeq{bthetab1} then immediately yields our estimate for $b_\theta^{bc}$,
\ba
b_{\theta}^{bc}(z) =\:& [(1+z)H_0]^{-1} \frac{\partial\delta_{\rm crit}(z)}{\partial(\theta_{bc,0}/H_0)} (1-b_1) \vs
\approx\:& 6.8 [(1+z)H_0]^{-1} (b_1-1)\,.
\label{eq:btheta}
\ea  
Clearly, this is of the same order of magnitude as estimated using the
simple local bias ansatz in \refsec{thresh}.  Note however the
different physics encoded in either approach: \refsec{thresh} considered
the case when the observed galaxy density depends on the baryon and CDM
densities individually, as is expected since the stellar mass of a galaxy
depends on the overall amount of baryons available, while the gravitational
collapse of halos depends on the total mass.  On the other hand, 
the estimate derived in \refeq{btheta} is based on the purely gravitational
effect of the decaying relative-velocity mode on the collapse of halos.  
Thus, unlike the effect discussed in \refsec{thresh}, this effect is
already present for halos.  The fact that both estimates agree at
the order-of-magnitude level further suggests that an estimate
of $b_\theta^{bc} \sim H_0^{-1}$ is robust.  
\refeq{btheta} is the fiducial value we will assume for
our results in \refsec{lin} and \refsec{1loop}.  

\subsection{Induced bias from streaming velocity}
\label{sec:bthetabv2}

The estimates for $b_\theta^{bc}$ provided so far model the bulk flow effect
of a relative velocity divergence.  Ref.~\cite{blazek/etal:15} argued
for a $b_\theta^{bc}$ induced by the relative velocity effect $\propto \vv_{bc}^2$ on very low-mass halos.  Specifically, they derived
\ba
b_\theta^{\rm B16} =\:& \frac23 b_{v^2} \, L_s(z)\, \s_{v_{bc}}^{-1}(z) \vs
\approx\:& 52 H_0^{-1} \left(\frac{b_{v^2}}{0.01 \s_{v_{bc}}^{-2}}\right) \left( \frac{D(z)}{D(1.2)} \frac{2.2}{1+z} \right)
\label{eq:bthetaBlazek}
\ea
where
\ba
L_s(z) \equiv\:& \s_{v_{bc}}^{-1}(z) \left\< v_{bc}^i \frac{\partial_i}{\nabla^2} \d \right\>
= \s_{v_{bc}}^{-1}(z) \int_{\vk} k^{-2} P_{\d\theta_{bc}}(k,z) \vs
\approx\:& 5.34 \Mpch \left(\frac{D(z)}{D(1.2)}\right)\,.
\label{eq:Ls}
\ea
Here, we have introduced $\int_{\vk} \equiv \int d^3\vk/(2\pi)^3$ and normalized to $z=1.2$.  This result was taken from a perturbation theory loop integral (see \refsec{1loop}).  At $z=1.2$ and for a fiducial value of $b_{v^2}=0.01$, this is roughly 8 times larger than the bias predicted from the excursion set \refeq{btheta} for $b_1=2$.  Note also the very different redshift scalings, where the prediction of \refeq{btheta} scales as $b_\theta \propto b_1(z) (1+z)^{-1}$, while \refeq{bthetaBlazek} scales as $D(z)/(1+z)$.  
   
Crucially, as discussed in \refsec{1loop} below, the loop integral leading to $b_\theta^{\rm B16}$ 
runs over very small-scale modes that are not modeled correctly in perturbation theory.  In a consistent perturbative expansion, it is 
absorbed by a renormalized bias parameter $b_\theta$, whose value is not 
calculable within perturbation theory and must be estimated using models
of galaxy and halo formation (such as the excursion set).  Still, it is 
reasonable to
wonder whether there is in fact a physical contribution to $b_\theta$ that
comes from the relative velocity effect $\propto v_{bc}^2$, and is thus
related to $b_{v^2}$.  Such a contribution would be physically independent
from, and presumably additive to, the bulk velocity divergence effect estimated using the 
spherical collapse calculation above.  We now derive which properties this
contribution would have to have.  

Consider a galaxy sample whose
local number density $n_g$ depends, at a given time $\tau$ and among various other quantities, on $v_{bc}^2$.  Specifically, we allow for $n_g(\vx)$
to depend on $v_{bc}^2$ in a finite region around $\vx$:
\be
n_g(\vx) = \int d^3 \vy\:\cdots F_g[ \vv_{bc}^2(\vx+\vy), \cdots ; \vy ]\,.
\label{eq:Fg1}
\ee
Here, the second ellipsis stands for other terms such as $\d_m(\vy')$, which each
come with an associated convolution integral $\int d^3 \vy'$, as denoted by the first ellipsis.  
We are interested in the linear response of $n_g$ to a long-wavelength perturbation
in the relative velocity, $\vv_{bc}^\ell$.  Hence, we write
\be
\vv_{bc} = \vv_{bc}^\ell + \vv_{bc}^s\,,
\ee
separating the long- and short-wavelength pieces.  At leading order,
$\vv_{bc}^s$ is uncorrelated with $\vv_{bc}^\ell$.  Further, we are interested
in long-wavelength perturbations that are much larger than the scale $R_*$ over
which galaxies form (for dark matter halos, this scale is the Lagrangian radius $R_* = R_L(M)$).  In the present context, $R_*$ is the typical
extent of the support of the functional kernel $F_g$ in \refeq{Fg1}.  
Then, we can perform a Taylor series up to second order in $\vv_{bc}$ to obtain
\ba
& n_g(\vx)\Big|_{\vv_{bc}^\ell} - n_g(\vx)\Big|_{0} \vs
& = \int d^3 \vy\: \left(\frac{\partial}{\partial \vv_{bc}^2} F_g[ 0, \cdots;\vy ] \right) 2 \vv_{bc}^s(\vx+\vy) \cdot \vv_{bc}^\ell(\vx+\vy) \vs
& \qquad + \int d^3 \vy\: \left(\frac{\partial}{\partial \vv_{bc}^2} F_g[ 0, \cdots;\vy ] \right) \left[\vv_{bc}^\ell(\vx+\vy) \right]^2
\vs
&=  \int d^3 \vy\: \left(\frac{\partial}{\partial \vv_{bc}^2} F_g[ 0, \cdots;\vy ] \right) \vv_{bc}^s(\vx+\vy) \cdot \vy \times \theta_{bc}^\ell(\vx) \vs
& \qquad + \int d^3 \vy\: \left(\frac{\partial}{\partial \vv_{bc}^2} F_g[ 0, \cdots;\vy ] \right) \times (\vv_{bc}^\ell)^2 (\vx) \vs
& \qquad + \O([\vv_{bc}^\ell]^3, \nabla^2 \theta_{bc}^\ell) 
\,.
\label{eq:Fg2}
\ea
In the second line, we have used
\be
(v_{bc}^\ell)^i (\vx+\vy) = (v_{bc}^\ell)^i (\vx) + y^k \partial_k(v_{bc}^\ell)^i (\vx) + \cdots\,.
\ee
The first term vanishes at linear order when inserted into the functional, while
of the second term $\propto \partial_k (v_{bc}^\ell)^i$ only the trace part,
namely $\theta_{bc}^\ell$ remains, in both cases due to the absence of
preferred directions in the small-scale modes.  We have dropped higher
derivative terms, which we will justify below.  We clearly obtain 
a bias with respect to $\vv_{bc}^2$, given by the ensemble average of 
the second line of \refeq{Fg2} over small-scale modes while keeping
large-scale modes fixed:
\ba
b_{v^2}^{bc} = \frac1{\bar n_g} \left\< \int d^3 \vy\: \cdots\left(\frac{\partial}{\partial \vv_{bc}^2} F_g[ 0, \cdots;\vy ] \right) \right\>\,.
\label{eq:bvsq1}
\ea
Here, we have set $\vx\to 0$, since the expectation value is independent of
position.  We also no longer need to explicitly restrict to the small-scale
component of $\vv_{bc}$.  
In addition, the first line of \refeq{Fg2} corresponds to
an effective bias w.r.t $\theta_{bc}$, which we denote as $b_\theta^{bc, v^2}$,
which is analogously given by
\ba
b_\theta^{bc, v^2} = \frac1{\bar n_g} \left\< \int d^3 \vy\:\cdots \left(\frac{\partial}{\partial \vv_{bc}^2} F_g[ 0, \cdots;\vy ] \right) \vv_{bc}(\vy) \cdot \vy \right\>\,.
\nonumber
\ea
Note that to obtain a nonzero value, we need to include the
dependence of $F_g$ on at least one other statistical field (as the expectation
value of any integral over $\vv_{bc}$ with a fixed kernel vanishes).  The leading expression is obtained by including a dependence on the local matter density perturbation $\d_m$,
$F_g \to F_g[\vv_{bc}^2, \d_m; \vy,\vy']$.  \refeq{Fg1} now becomes a functional
in both $\vv_{bc}^2$ and $\d_m$, and the leading term is
\ba
b_\theta^{bc, v^2} =\:& \frac1{\bar n_g} \int d^3 \vy \int d^3 \vy'\: \left(\frac{\partial^2}{\partial \vv_{bc}^2 \partial\d_m} F_g[ 0, 0;\vy,\vy' ] \right) \vs
& \hspace*{2cm}\times\left\<  \d_m(\vy') \vv_{bc}(\vy) \cdot \vy \right\> \vs
\sim\:& b_{v^2}^{bc}\,\left\<  \d_m\, \vv_{bc} \cdot \vy \right\>_{F_g} \,.
\label{eq:bthv2}
\ea
In the second line of \refeq{bthv2}, we have used that the prefactor is expected to be 
of the same order of magnitude as $b_{v^2}^{bc}$ given in \refeq{bvsq1}, while the
expectation value is essentially a generalized second moment (with window function normalized to unity) of $\d_m$ and
$\vv_{bc}\cdot \vy$ on the scale $R_*$.   
The physical interpretation of this expression is the following.  
The number density of observed galaxies depends on the distribution 
of $v_{bc}^2(\vy)$ within a volume of scale $R_*$, for
example via the effect $\vv_{bc}^2$ has on the abundance of very low-mass
halos.  The bias $b_{v^2}^{bc}$ corresponds to the volume average
of this dependence (the fact that $b_{v^2}^{bc}$ quantifies an average over very small-scale effects at high redshifts is the reason for the significant uncertainty in its magnitude).  Now, a long-wavelength relative velocity divergence $\theta_{bc}$ 
induces a nontrivial radial profile in $v_{bc}^2$.  This profile depends
on the small-scale contributions to $v_{bc}$, and thus averages to zero
in the absence of other perturbations.  However, the small-scale relative
velocity correlates with small-scale density perturbations, i.e. regions within the volume $\sim R_*^3$ considered
with large density contrast also have larger amplitudes of the relative
velocity on average, and this leads to a net effect of the induced 
relative velocity profile and hence $\theta_{bc}$ on the galaxy abundance.  

In order to obtain a quantitative estimate, we work to lowest order in perturbation theory, where $\d_m,\,\vv_{bc}$ are Gaussian fields.  Then, the moment
in the last line of \refeq{bthv2} can always be written as
\be
\left\<  \d_m\, \vv_{bc} \cdot \vy \right\>_{F_g} = R_* \int_0^\infty \frac{k^2 dk}{2\pi^2}
\,W_{F_g}(k) k^{-1} P_{\d\theta_{bc}}(k)\,,
\label{eq:Fgmom}
\ee
where $W_{F_g}(k)$ is dimensionless and we have pulled out a factor $R_*$ since by assumption $|\vy|$ is of order $R_*$.  If $F_g$ is a smooth function as expected physically, then $W_{F_g}(k)$ will drop off quickly for $k \gg 1/R_*$.  Furthermore, low-$k$ modes with $k \ll 1/R_*$ also cannot contribute to \refeq{Fgmom}, since the angle average over $\vv_{bc}\cdot\vy$ for a constant $\vv_{bc}$ vanishes.  Thus, in the limit $k R_* \ll 1$, the kernel has to scale as $W_{F_g}(k) = \O(k^2 R_*^2)$; this is easily verified with concrete example kernels in \refeq{bthv2}.  This means that only modes with $k$ of order $R_*$ will contribute appreciably to \refeq{Fgmom}.  Note that this is completely generic, and follows directly from the derivative expansion in \refeq{Fg2}.  That is, any small-scale modes of $\vv_{bc}$ that contribute to the physical bias parameter $b_\theta^{bc}$ cannot be of much larger scale than $R_*$.  This continues to hold if one continues the expansion in derivatives in \refeq{Fg2} to higher order (yielding terms of order $R_*^{2n}\partial^{2n} \theta_{bc}$), and is another qualitative difference to \refeqs{bthetaBlazek}{Ls}.

\begin{figure}[t!]
\centering
\includegraphics[width=0.49\textwidth]{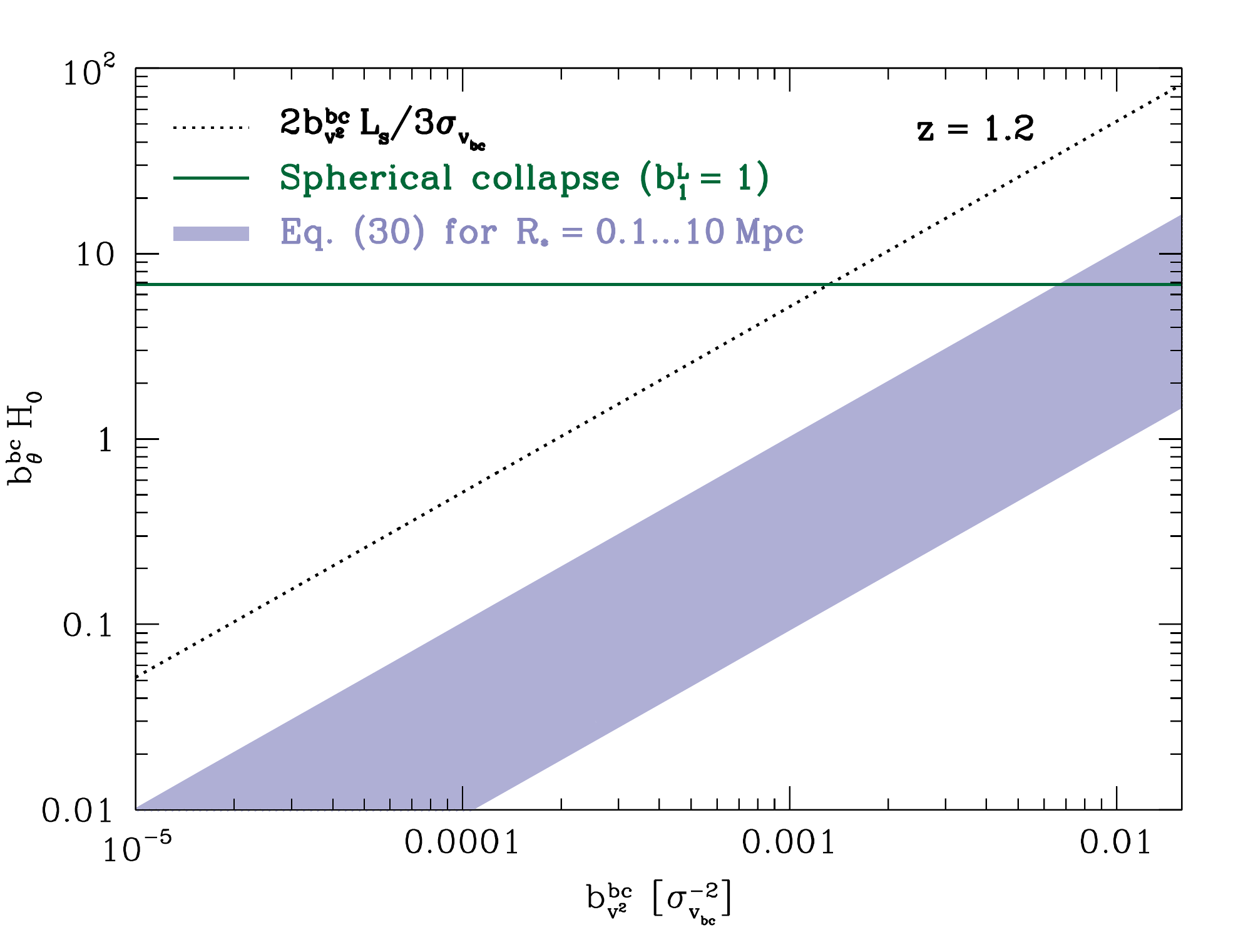}
\caption{Various estimates of the bias parameter $b_\theta^{bc}$ as a function of $b_{v^2}^{bc}$.  The solid line shows the excursion set estimate \refeq{btheta} from \refsec{scollapse}, which is independent of $b_{v^2}^{bc}$.  The shaded band shows the rough range of values expected for the induced bias from streaming velocity effects on small scales [\refeq{bthv2e}].  The dotted line shows the loop integral derived in \cite{blazek/etal:15} [\refeq{bthetaBlazek}] which is argued to be unphysical. 
\label{fig:btheta}}
\end{figure}
In summary, we obtain as an order-of-magnitude estimate 
\ba
b_\theta^{bc, v^2} \sim\:& b_{v^2}^{bc} \left(\frac{ k^2 P_{\d\theta_{bc}}(k)}{2\pi^2}\right)_{k\sim\pi/R_*} \,R_* \label{eq:bthv2e}\\
\sim\:& 6 H_0^{-1} \left(\frac{b_{v^2}}{0.01\,\s_{v_{bc}}^{-2}}\right) 
\  \mbox{for}\    R_* \approx 5\Mpch,\,z=1.2
\,.
\nonumber
\ea
\reffig{btheta} (shaded band) shows the estimate for $b_\theta^{bc,v^2}$ as a function of
$b_{v^2}^{bc}$ for a range of spatial scales $R_* = 0.1-10\Mpch$.  Since $k^2 P_{\d\theta_{bc}}(k)$ shows significant oscillations, we average it over $k$ with a Gaussian kernel centered on $\pi/R_*$ and with 1D RMS of $2\pi/R_*$, rather than taking the value at exactly $\pi/R_*$.  In any case, our estimate of $b_{\theta}^{bc,v^2}$ can only be taken as a rough approximation.  \reffig{btheta} also shows the estimate \refeq{btheta} from the previous section, and the result from the loop integral \refeq{bthetaBlazek}.  
Clearly, the latter is much larger than either of the two physical estimates if $b_{v^2}^{bc} \gtrsim 10^{-3}$.  
The reason is that the integral in \refeq{Ls} extends over all modes
(in particular large-scale modes with $k \ll 1/R_*$), while we have argued that physically, only modes with $k$ around $1/R_*$ should contribute.  
Whether the contribution $b_\theta^{bc,v^2}$ dominates over the bulk flow contribution \refeq{btheta} clearly depends on the value of $b_{v^2}^{bc}$;  however, for most of the open parameter space of $b_{v^2}^{bc}$ the contribution from \refeq{btheta} is larger.

\begin{figure}[t!]
\centering
\includegraphics[width=0.49\textwidth]{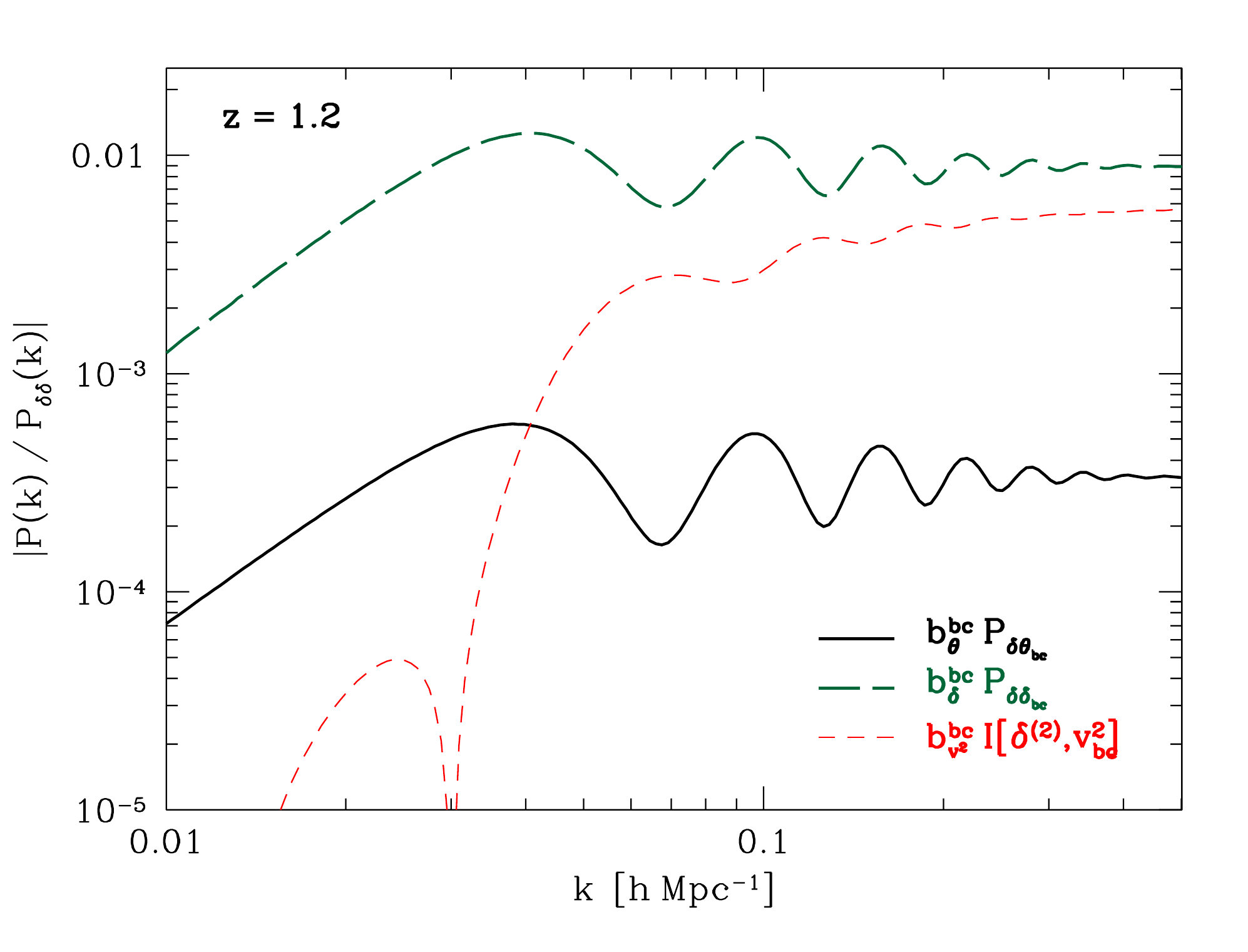}
\vspace*{-0.5cm}
\caption{Leading baryon-CDM perturbation contributions to the galaxy power spectrum (absolute magnitude divided by $b_1$ at $z=1.2$), relative to the linear matter power spectrum (solid: $\theta_{bc}$; long-dashed: $\d_{bc}$).  Here we have used $b_\delta^{bc}=1$ and $b_{\theta}^{bc} = 6.8/[(1+z) H_0]$ [\refeq{btheta}, setting $b_1=2$].  We also show the largest of the contributions from the term in \refeq{dgv2}, $b_{v^2}^{bc} \mathcal{I}^{[\d^{(2)}, v_{bc}^2]}(k)$, which enter at 1-loop order (short-dashed; \refsec{1loop}), 
assuming $b_{v^2}^{bc} = 0.01 \sigma_{v_{bc}}^{-2}$.  
\label{fig:Pklin}}
\end{figure}

\subsection{Galaxy power spectrum at linear order}
\label{sec:lin}

Using \refeq{dg2}, we can immediately write down the contributions from
baryon-CDM perturbations to the linear galaxy auto and galaxy-matter cross power spectrum:
\ba
P_{gg}^{\rm lin}(k)\Big|_{bc} =\:& 2 b_1 b_\theta^{bc} P_{\d\theta_{bc}}(k) 
+ 2 b_1 b_\delta^{bc} P_{\d\delta_{bc}}(k) \vs
& + (b_\theta^{bc})^2 P_{\theta_{bc}\theta_{bc}}(k) 
+ (b_\delta^{bc})^2 P_{\delta_{bc}\delta_{bc}}(k) \vs
& + 2 b_\delta^{bc} b_\theta^{bc} P_{\delta_{bc}\theta_{bc}}(k) 
\vs
P_{gm}^{\rm lin}(k)\Big|_{bc} =\:& b_\theta^{bc} P_{\d\theta_{bc}}(k) 
+ b_\delta^{bc} P_{\d\delta_{bc}}(k) 
\,,
\label{eq:Plin}
\ea
where $P_{xy}(k)$ denote linear cross-power spectra between the matter
density ($\d$), the relative density perturbation ($\d_{bc}=R_+$) and
the relative velocity divergence ($\theta_{bc} = R_- H_0$).  \reffig{Pklin}
shows the leading contributions, i.e. those involving only one power
of $\d_{bc},\,\theta_{bc}$, divided by $b_1$ and the linear
matter power spectrum $P_{\d\d}(k)$, at $z=1.2$.  We use the transfer
function output of CAMB \cite{camb}, from which we obtain $R_+,\,R_-$
via the matching described in \refapp{Tk}.  Note that both
$\d_{bc}$ and $\theta_{bc}$ contributions have a very similar scale dependence.  
Clearly, despite 
the large value of $b_{\theta}^{bc}$ as compared to $b_\delta^{bc}$, the
baryon-CDM density perturbation $\d_{bc}$ is by far the largest contribution,
being roughly scale independent for $k\gtrsim 0.05\iMpch$ at the level of $\sim 1\%$.  
Given that the correlation coefficient of all the fields $\d,\,\d_{bc},\,\theta_{bc}$ is unity, we also infer immediately that the contributions in \refeq{Plin} involving two powers
of $\d_{bc},\,\theta_{bc}$ are highly suppressed, and contribute at most
at the $\sim 0.01\%$ level.

The operator $v_{bc}^2$ considered in the previous literature, being quadratic,
only contributes to $P_{gg}(k),\,P_{gm}(k)$ at the 1-loop level.  
We will turn to this in \refsec{1loop}.  \reffig{Pklin} also shows the largest 
of the 1-loop contributions $\propto v_{bc}^2$, assuming a value $b_{v^2}^{bc}=0.01\sigma_{v_{bc}}^{-2}$ at the upper end of the expected range.  
It is clearly larger than the contribution from $\theta_{bc}$, but smaller than
that from $\d_{bc}$.  While these conclusions depend on the values of the
various bias parameters, we generically expect $\d_{bc}$ to provide the
largest baryon-CDM contribution to the galaxy power spectrum.  
We will discuss the significance of these contributions for the BAO
standard ruler in \refsec{concl}.  

\section{Galaxy clustering and baryon-CDM perturbations beyond linear order}
\label{sec:nonlin}

We now generalize the results of the previous section to nonlinear order
in perturbation theory.  We begin by deriving all operators that
have to be included in the bias expansion \refeq{dggen} of a general galaxy sample
in the presence of baryon-CDM perturbations, that is for nonzero $\d_{bc},\,\theta_{bc},\,\vv_{bc}$.  
We then provide the complete description of the 1-loop galaxy power spectrum,
i.e. the leading nonlinear correction to the results of \refsec{lin}.  

\subsection{General bias expansion}
\label{sec:biasgen}

The complete bias
expansion in the case of Gaussian initial conditions and perturbations
that are exclusively in the adiabatic growing mode has recently been 
derived \cite{senatore:2014,MSZ}.  We will build on those results.  
To start, let us derive the
equations for the nonlinear system of the coupled baryon-CDM fluids,
i.e. the nonlinear generalization of \refeq{deltalinearEOM1}.  
Denoting  the velocity of the total matter fluid as 
$v_m^i = f_b v_b^i + (1-f_b) v_c^i$ and
defining the convective time derivative 
\be
\frac{D}{D\tau} \equiv \frac{\partial}{\partial\tau} + v_m^i \frac{\partial}{\partial x^i}\,,
\label{eq:DDtau}
\ee
we obtain (see also App.~C in \cite{blazek/etal:15} and \cite{ahn} for the Eulerian expressions)
\ba
\frac{D}{D\tau} \d_s +\theta_s  =\:& - \d_s \theta_s 
- g_s v_r^i \partial_i \d_s \label{eq:deltanlEOM}\\
\left( \frac{D}{D\tau} + \cH \right) \theta_s + \frac32 \Om \cH^2 \d_m =\:& 
- (\partial^i v_s^k)^2 - g_s v_r^i \partial_i \theta_s\,,
\nonumber
\ea
where $s=b,c$ and we have introduced the shorthand $g_b = 1-f_b$ and $g_c = -f_b$.  
On the l.h.s. we now have the standard differential operator
for the gravitational evolution, while the r.h.s. contains the nonlinear
terms due to gravitational evolution and the relative velocity.  
It is clear that $\d_m,\theta_m$ and the relative density and velocity
no longer decouple at nonlinear order.  
Noting that $D/D\tau$ is invariant under homogeneous but time-dependent
translations (boosts), these equations are explicitly boost-invariant.  
That is, all quantities that appear in \refeq{deltanlEOM} are 
local observables (note in particular the absence of $\vv_m$).  

\refeq{deltanlEOM} gives us a strong hint as to which operaters we 
should allow to appear in the bias expansion in the two-fluid system.  
We have to include $\vv_{bc}$, $\d_b$, $\d_c$ as
well as $\theta_b,\,\theta_c$ and $\partial^i v_b^j$, $\partial^i v_c^j$
in the bias expansion (see also Sec.~3.1 in \cite{angulo/etal:2015}).  Again, it is convenient to decompose these in
terms of the adiabatic growing mode $\d_m,\,\partial^i v_m^j$ and
the baryon-CDM perturbations $R_+,\,R_-$.  The latter modes are captured
in the bias expansion by including
\be
R_+(\vq)\quad\mbox{and}\quad R_-^i \equiv \frac{\partial_q^i}{\nabla^2_q} R_-(\vq)\,, 
\label{eq:Rplusminus}
\ee
and higher spatial derivatives of these quantities (but no time derivatives, see below).  Here $\vq$ is the Lagrangian coordinate.  Evaluating $R_\pm$ at the Lagrangian position is in fact required in order to be able to renormalize the operators consistently \cite{assassi/baumann/schmidt}. 
Before proceeding to the general bias expansion, we should consider the
importance of spatial derivatives.  From \refeq{deltanlEOM} it is clear that
$R_+$ and $\partial^i R_-^j$ have the same status as $\d_m,\,\partial^i v_m^j$.  
These terms are thus to be considered as lowest order in derivatives.  
Higher spatial derivatives, for example $\partial^k R_+$ or $\nabla^2 R_-^i$
are then expected to be suppressed by the same scale that appears in higher
derivative operators of the adiabatic quantities, for example $\partial^k \d_m$.  This is the spatial scale $R_*$, which for halos is of order the Lagrangian radius, or smaller.  

Let us now derive the complete set of operators of the bias expansion
at lowest order in derivatives.  Formally, this is what is obtained in the
limit $R_* \to 0$.  We begin with a recap of the adiabatic growing mode
case.  The basis of operators in this case can be conveniently constructed
out of the tidal tensor $\Pi^{[1]}_{ij} \equiv 2\partial_i\partial_j\Phi/(3\Om \cH)^2$, which contains the density (trace part) and tidal field.  
Here, a superscript $[n]$ denotes operators that {\it start} at $n$-th order in perturbation theory, while $n$-th order {\it contributions} to
an operator are denoted with a superscript $(n)$.  
In addition, we have to include convective time derivatives of $\Pi^{[1]}_{ij}$.  
We then define
\be
\Pi^{[n]}_{ij} \equiv \frac{1}{(n-1)!} \left[(\cH f)^{-1}\frac{D}{{D}\tau} \Pi^{[n-1]}_{ij} - (n-1) \Pi^{[n-1]}_{ij}\right] \,,
\ee
which by construction is an operator that starts at $n$-th order in perturbation theory.  The basis then consists of all scalar combinations of the 
$\Pi^{[n]}_{ij}$ up to the desired perturbative order, with the exception of ${\rm Tr}[\Pi^{[n]}]$, $n>1$, which is expressible in terms of lower order operators \cite{MSZ}.  For example, up to third order, we then have the following list of bias operators for
Gaussian initial conditions~\cite{MSZ}: 
\bea
{\rm 1^{st}} \ && \ {\rm Tr}[\Pi^{[1]}] 
\label{eq:listP} \\[3pt] 
{\rm 2^{nd}} \ && \ {\rm Tr}[(\Pi^{[1]})^2]\,,\  ({\rm Tr}[\Pi^{[1]}])^2 \nonumber\\[3pt] 
{\rm 3^{rd}} \ && \ {\rm Tr}[(\Pi^{[1]})^3 ]\,,\ {\rm Tr}[(\Pi^{[1]})^2] \hskip 1pt  {\rm Tr}[\Pi^{[1]}]\,,\ ({\rm Tr}[\Pi^{[1]}])^3\,,\vs
&& \  {\rm Tr}[\Pi^{[1]} \Pi^{[2]}]\,, \nonumber
\eea
where all operators are evaluated at the same Eulerian position and time $(\vx,\tau)$.  These are simply all scalar combinations of the matter density and tidal field, with the exception of the last term in the third line which involves a time derivative and is, at a fixed time, nonlocally related to the density
and tidal field \cite{senatore:2014,MSZ}.  

We now simply augment the list \refeq{listP} by all scalar combinations of
$\Pi^{[n]}_{ij}$ with $R_+,\,R_-^i,\,\partial^j R_-^i$.  
Unlike for the quantities $\d_m,\,\partial^i v_m^j$ which
grow under gravity, we do not have to consider time derivatives of $R_+$ and $R_-^i$ in the bias expansion.  
To see this, we first allow  the galaxy density at a given
point to depend on $\d_r$, $\vv_r$ along the past fluid
trajectory leading to this point.  Up to any order in perturbation theory, this relative velocity along the fluid trajectory can be written as a local function of $R_+,\,\v{R}_-$ and the $\Pi^{[n]}$ ($n \geq 1$) appearing in the galaxy bias expansion.  There are corrections from the relative displacement of the baryon and CDM fluids, encoded in the terms $v_r^i\partial_i \d_s,\,v_r^i\partial_i\theta_s$
in \refeq{deltanlEOM}; however, these lead to higher derivative terms which we will discuss below.  
Thus, by allowing for $R_+,\,,\v{R}_-$ to appear in all combinations with the $\Pi^{[n]}$, we allow for a general dependence of the galaxy density on baryon-CDM relative density and velocity perturbations along the past fluid trajectory.
This is similar to the case of non-Gaussian initial conditions, where an additional field $\phi(\vq)$ appears in the bias expansion, again without time derivatives \cite{assassi/baumann/schmidt}.  

Note that, once all terms at a given order are included, we can equivalently
consider the derivatives acting on $R_-$ [\refeq{Rplusminus}] as being
with respect to Eulerian coordinate $\vx$, since the Jacobian $\partial \vx^i/\partial\vq^j$ is expressible in terms of the other terms included
in the general bias expansion.  
Defining $\v{R}_- = R_-^i$, and adopting matrix notation, where 
$(\nabla R_-)^{ij} \equiv \partial^i R_-^j$, we have to augment the list \refeq{listP} up to cubic order by
\bea
{\rm 1^{st}} && \ R_+(\vq)\,,\  {\rm Tr}[\nabla R_-](\vq) 
\label{eq:listR} \\[3pt] 
{\rm 2^{nd}} && \ R_+ {\rm Tr}[\Pi^{[1]}]\,,\  (\v{R}_-)^2\,,\  {\rm Tr}[(\nabla R_-) \Pi^{[1]} ] \,,\vs
&& {\rm Tr}[\nabla R_-] {\rm Tr}[\Pi^{[1]}]
\nonumber\\[3pt] 
{\rm 3^{rd}} && R_+ {\rm Tr}[(\Pi^{[1]})^2]\,,\  R_+ ({\rm Tr}[\Pi^{[1]}])^2 \,,\  (\v{R}_-)^2 {\rm Tr}[\Pi^{[1]}]\,,\vs
&& \v{R}_- \Pi^{[1]} \v{R}_-\,,\  {\rm Tr}[(\nabla R_-) (\Pi^{[1]})^2]\,,\vs
&& {\rm Tr}[(\nabla R_-) \Pi^{[1]}]{\rm Tr}[\Pi^{[1]}]\,, \  
{\rm Tr}[\nabla R_-] {\rm Tr}[(\Pi^{[1]})^2]\,,\vs
&& {\rm Tr}[\nabla R_-] ({\rm Tr}[\Pi^{[1]}])^2\,,\  {\rm Tr}[(\nabla R_-) \Pi^{[2]}]
\,, \nonumber
\eea
where $R_\pm$ are all evaluated at the Lagrangian position $\vq[\vx,\tau]$
while $\Pi^{[n]}_{ij}$ are evaluated at $(\vx,\tau)$.  
Here, we have restricted to terms that are linear in $R_\pm$.  
However, we have kept terms involving $(R_-^i)^2 \propto v_{bc}^2$.  The reasoning is that there is possibly a hierarchy between $b_{v^2}^{bc}$ and $b_\d^{bc},\,b_\theta^{bc}$ as we have seen in \refsec{biaslin}.  \reffig{Pklin} shows
that the terms that are higher order in $R_\pm$ will be highly suppressed (although 
it is straightforward to include them).  
Further, following our discussion about higher derivatives, we neglect the term $R_-^i\partial_i\d$, since its bias parameter is expected to be of order $R_*$, whereas the coefficient multiplying $(\partial_i R_-^i) \d$ is of order $H_0^{-1}$  [\refeq{btheta}].  This type of term contains the effects of advection of the baryon fluid with respect to the CDM fluid.  Note that the relative displacement between baryons and CDM generated by the relative velocity, that is the integral $\int d\tau\,\v{v}_{bc}$ over the advection terms on the right-hand side of \refeq{deltanlEOM}, is very small, of order of kpc.  In any case, terms of this type are straightforward to
include (see \cite{lewandowski/perko/senatore}, who perform a resummation of 
this type of term at higher orders).  

We can equivalently express the list in \refeq{listR} in a slightly more
familiar form, in terms of the matter density 
and tidal field $K_{ij} \equiv (\partial_i\partial_j/\nabla^2 - \d_{ij}/3) \d_m$, 
\bea
1^{\rm st}\  && \d_{bc},\,\theta_{bc} \vs
2^{\rm nd}\  && \d_{bc} \d,\, v_{bc}^2,\,\theta_{bc}\d,\, K_{ij} \partial^i v_{bc}^j \vs
3^{\rm d}\  && \d_{bc}\d^2,\, \d_{bc} (K_{ij})^2,\, v_{bc}^2 \d,\, K_{ij} v_{bc}^i v_{bc}^j,\,\theta_{bc}\d^2,\, \theta_{bc} (K_{ij})^2, \vs 
&& \d K_{ij} \partial^i v_{bc}^j,\, K_{ij} K^j_{\  k} \partial^i v_{bc}^k,\,O_{\rm nloc}^{bc}\,,
\label{eq:listR2}
\eea
where from now on we let $\d \equiv \d_m$ for clarity, and
\ba
O_{\rm nloc}^{bc} \equiv \frac{8}{21} \left(\partial^i v_{bc}^j -\frac13 \d^{ij}\theta_{bc}\right)
\frac{\partial_i\partial_j}{\nabla^2}
\left[ \d^2 - \frac32 (K_{lm})^2 \right]\,.
\nonumber
\ea
Up to other cubic terms already included in \refeq{listR2}, $O_{\rm nloc}^{bc}$ can also be written as $(4/5)\,{\rm Tr}[R_- \Pi^{[2]}]$.  It is clearly nonlocally related to $\d,\,K_{ij}$.  This is
the first instance of a convective time derivative appearing in the bias
expansion.  Equivalently, it can be seen as a generalization of $\Gamma_3$ defined in \cite{assassi/etal}.  

Apart from $v_{bc}^2$ \cite{tseliakhovich/hirata:2010,dalal/etal:2010}, $\d_{bc}$ \cite{barkana/loeb:11},
and $\theta_{bc},\, \d v_{bc}^2,\, K_{ij} v_{bc}^i v_{bc}^j$ \cite{blazek/etal:15}, all terms in \refeqs{listR}{listR2} are introduced here for the first time.  
In general, each of these terms is associated with a respective bias parameter that is specific to any given galaxy sample.  However, many of the terms are linked by similar physics, so we can estimate the order of magnitude of the bias coefficients that are associated to each operator: 

\emph{(i) operators involving $R_+ = \d_{bc}$:}  the coefficients are expected to be of order one, that is of the same order as the ordinary, growing-mode bias parameters multiplying $\d,\,(K_{ij})^2,\cdots$.  Whether there is an enhancement for rare massive halos like in the case of the density bias parameters is unclear however.  

\emph{(ii) operators involving $v_{bc}^i v_{bc}^j$:}  the coefficients of these terms are expected to be of order $b_{v^2}$, which as discussed in \refsec{intro} could be as large as $0.01 \sigma_{v_{bc}}^{-2}$, or as small as $10^{-5} \sigma_{v_{bc}}^{-2}$.  

\emph{(iii) operators involving $\partial^i v_{bc}^j$:}  these include $\theta_{bc}$, and we expect the associated bias parameters to be of similar order as $b_\theta^{bc} \sim {\rm few}\, H_0^{-1}$.  

As we will see in \refsec{1loop}, the hierarchy between the bias parameters determines which terms are the most relevant in the prediction for galaxy statistics.  The potential large magnitude of $b_{v^2}^{bc}$ compared to what is expected for the other two classes of terms was already used in \refeqs{listR}{listR2}, where we keep terms with two powers of $v_{bc}^i$ but only one power of $\d_{bc},\,\partial^i v_{bc}^j$.  

In addition to the deterministic operators listed in \refeqs{listR}{listR2}, there are also stochastic contributions induced by integrating out the small-scale perturbations;  physically, two galaxies in the same large-scale environment form from different realizations of the small-scale density field and their properties will correspondingly scatter around the expectation value.  This can be taken into account consistently in the bias expansion by introducing stochastic fields $\eps_O$ for each operator in the list \refeqs{listP}{listR} \cite{MSZ}, where $\eps_O$ have zero mean and are at lowest order in derivatives fully characterized by their one-point moments.  Note that for an $n$-th order operator $O^{[n]}$, these terms are order $n+1$ in perturbation theory.  Further, in case of the galaxy two-point function, these terms are fully captured by including ``contact terms'' $P_g(k) \supset N_0 + N_2 (R_* k)^2 + N_4 (R_*k)^4 +\cdots$ \cite{assassi/etal}, which are already present in the absence of baryon-CDM perturbations.  Thus, baryon-CDM perturbations do not add new stochastic contributions at the two-point function level.

Further, \refeqs{listR}{listR2} are only complete at lowest order in spatial derivatives.  The fact that galaxy formation is not perfectly local induces additional operators with higher spatial derivatives.  Essentially, we have to allow for any scalar combination of $\partial\cdots\partial \Pi^{[n]}_{ij}$, $\partial\cdots\partial R_+$, and $\partial\cdots\partial R_-^i$.  
Each derivative comes with the spatial scale $R_*$ of galaxy formation.  The two leading higher-derivative operators are
\be
R_*^2 \nabla^2 \d_{bc},\  R_*^2 \nabla^2 \theta_{bc}\,.
\label{eq:hderiv}
\ee
Note that we have already included all first order derivatives of $v_{bc}^i$ in \refeqs{listR}{listR2}, as they are not necessarily suppressed by $R_*$ (nevertheless, these terms could contain contributions induced by the nonlocality of galaxy formation, as described in \refsec{bthetabv2}).  
Higher-derivative terms generally become relevant on small scales.  For example, for values of $R_* \lesssim 10\Mpch$, the terms in \refeq{hderiv} are comparable to the next-to-leading contributions from higher order bias terms (see \refsec{1loop}).  

Finally, as noted by \cite{blazek/etal:15}, the fact that $R_\pm$ are evaluated at the Lagrangian position introduces further terms (again this is analogous to the case for primordial non-Gaussianity \cite{tellarini/etal,assassi/baumann/schmidt}),
\ba
2^{\rm nd}:\  & -s^k \partial_k \d_{bc},\,-s^k \partial_k \theta_{bc} \label{eq:listsO}\\
3^{\rm d}:\  & \frac12 s^l s^k \partial_l \partial_k \delta_{bc},\,
\frac12 s^l s^k \partial_l \partial_k \theta_{bc},\,
-s^k \partial_k v_{bc}^2 , \vs
& -\d \,s^k \partial_k \d_{bc},\,-\d \,s^k \partial_k \theta_{bc},\,-s^k K_{ij} \partial_k \partial^i v_{bc}^j
\,,
\nonumber
\ea
where $\v{s}$ is the displacement from the Lagrangian to Eulerian position, given by $s^i  = -\partial^i/\nabla^2 \d$ at linear order.  Each term in \refeq{listsO} is multiplied by the bias parameter of the corresponding leading operator in \refeq{listR2}.  They thus do not introduce additional free parameters.

\subsection{Galaxy power spectrum at 1-loop order}
\label{sec:1loop}

In order to derive the next-to-leading 1-loop contribution to the
galaxy power spectrum, we need to go to third order in perturbation theory,
and consequently need to consider all terms given in \refeq{listR2}.  
Throughout, we only keep terms at linear order in $\d_{bc},\,\theta_{bc}$,
as the higher order terms are highly suppressed (\refsec{lin}), 
but keep quadratic terms in $v_{bc}^2$; the extension to include all nonlinear
terms in $\d_{bc},\,\theta_{bc}$ is straightforward.  

Before presenting the calculation, we begin with some general considerations.  
The 1-loop contributions to $P_{gg},\,P_{gm}$ involve an integral $\int_{\vp} \equiv \int d^3\vp/(2\pi)^3$ over a loop momentum which can include contributions from very small scales.  Whether the integral converges to a finite value or not, these scales are not modeled physically by perturbation theory.  Instead, these contributions need to be isolated and removed, a procedure known from field theory as renormalization \cite{mcdonald/roy:2009,assassi/etal}.  This can be done by introducing an artificial cutoff $\Lambda$ and then adding counterterms to cancel the cutoff-dependent loop contributions.  The end result is that unphysical contributions are absorbed in renormalized bias parameters of lower order terms (or stochastic terms).  In the following we will describe this briefly in the context of the baryon-CDM contributions.  For this, we divide the terms in \refeq{listR2} into two classes.    

$(i)$ terms constructed out of $\d_{bc},\,\theta_{bc}$ and $\partial^i v_{bc}^j$ together with $\d$ and $K_{ij}$:  these have the same structure as those present
in standard perturbation theory, and we can use results of the latter by simply replacing
$P_{\d\d}(k)$ with $P_{\d\delta_{bc}}(k)$ or $P_{\d\theta_{bc}}(k)$, as appropriate, in the loop integrals.  
As shown in \cite{mcdonald/roy:2009,assassi/etal}, the only terms that remain after renormalization of the 1-loop galaxy power spectrum are
$\d^2,\  \G_2\equiv(K_{ij})^2-(2/3)\d^2$, and $O_{\rm nloc} \propto (4/5) {\rm Tr}[\Pi^{[1]}\Pi^{[2]}]$, so that the corresponding baryon-CDM terms are
\be
b^{bc}_{\d\d}\,[\delta_{bc} \d],\   
b^{bc}_{\theta\d}\,[\theta_{bc} \d],\   
b^{bc}_{\G}\,[\G^{bc}],\  
b^{bc}_{\rm nloc} [O_{\rm nloc}^{bc}]\,,
\nonumber
\ee
where brackets denote renormalized operators and
\ba
\G^{bc} \equiv\:& K_{ij} \partial^i v_{bc}^j -(2/3)\theta_{bc}\d 
\,.
\ea
None of these terms has been included in previous calculations
of baryon-CDM contributions to galaxy clustering \cite{dalal/etal:2010,yoo/dalal/seljak,yoo/seljak,slepian/eisenstein,blazek/etal:15}. 

$(ii)$ terms constructed out of $\v{v}_{bc}$ (without derivatives) and/or
$\v{s}$: these have a different structure, and need to be dealt with separately.  Apart from the term $v_{bc}^2$ considered in \cite{dalal/etal:2010,yoo/dalal/seljak,yoo/seljak,slepian/eisenstein,blazek/etal:15}, we also find new contributions from
$s^k \partial_k \theta_{bc}$ and $s^k\partial_k \d_{bc}$.  All other
terms either vanish by symmetry or renormalize $b_1$, $b_\theta^{bc}$, or higher derivative terms.  
Let us briefly consider the term $s^k \partial_k v_{bc}^2$, contributing to $P_{gg}^{\text{1-loop}},\, P_{gm}^{\text{1-loop}}$ through
\ba
& \< \d(\vk) (s^k \partial_k v_{bc}^2)(\vk') \> = 2 \< \d(\vk) \theta_{bc}(\vk') \> \int_{\vp} \frac{(\vk\cdot\vp)^2}{k^2 p^4} P_{\d\theta_{bc}}(\vp) \vs
&\qquad = \frac43 \< \d(\vk) \theta_{bc}(\vk') \> \int_{\vp} p^{-2} \, P_{\d\theta_{bc}}(\vp)\,.
\label{eq:dsdv}
\ea
This term is absorbed in the tree-level contribution via the renormalized bias coefficient $b_\theta^{bc}$, since it is given by $\< \d(\vk) \theta_{bc}(\vk')\>$ multiplied by a cutoff-dependent integral [see \refeqs{bthetaBlazek}{Ls}].  Within renormalized perturbation theory, there is no prediction for the physical bias $b_\theta^{bc}$, which can only be estimated through toy models of the small-scale physics of galaxy formation.  In \refsec{biaslin}, we have provided several estimates which are summarized in \reffig{btheta}.  These are significantly smaller than the fiducial value adopted by \cite{blazek/etal:15}.  

\begin{figure*}[t!]
\centering
\includegraphics[width=0.49\textwidth]{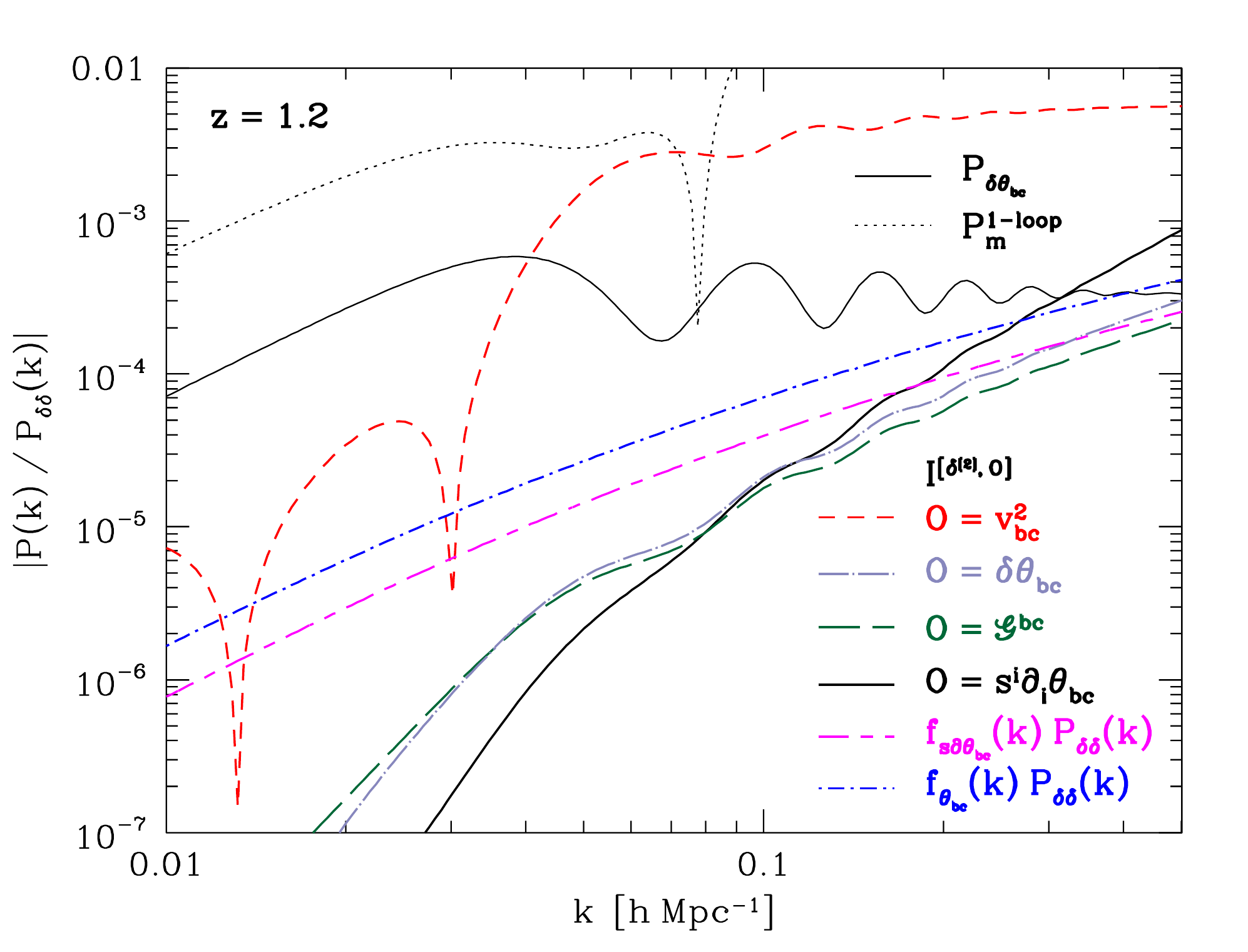}
\includegraphics[width=0.49\textwidth]{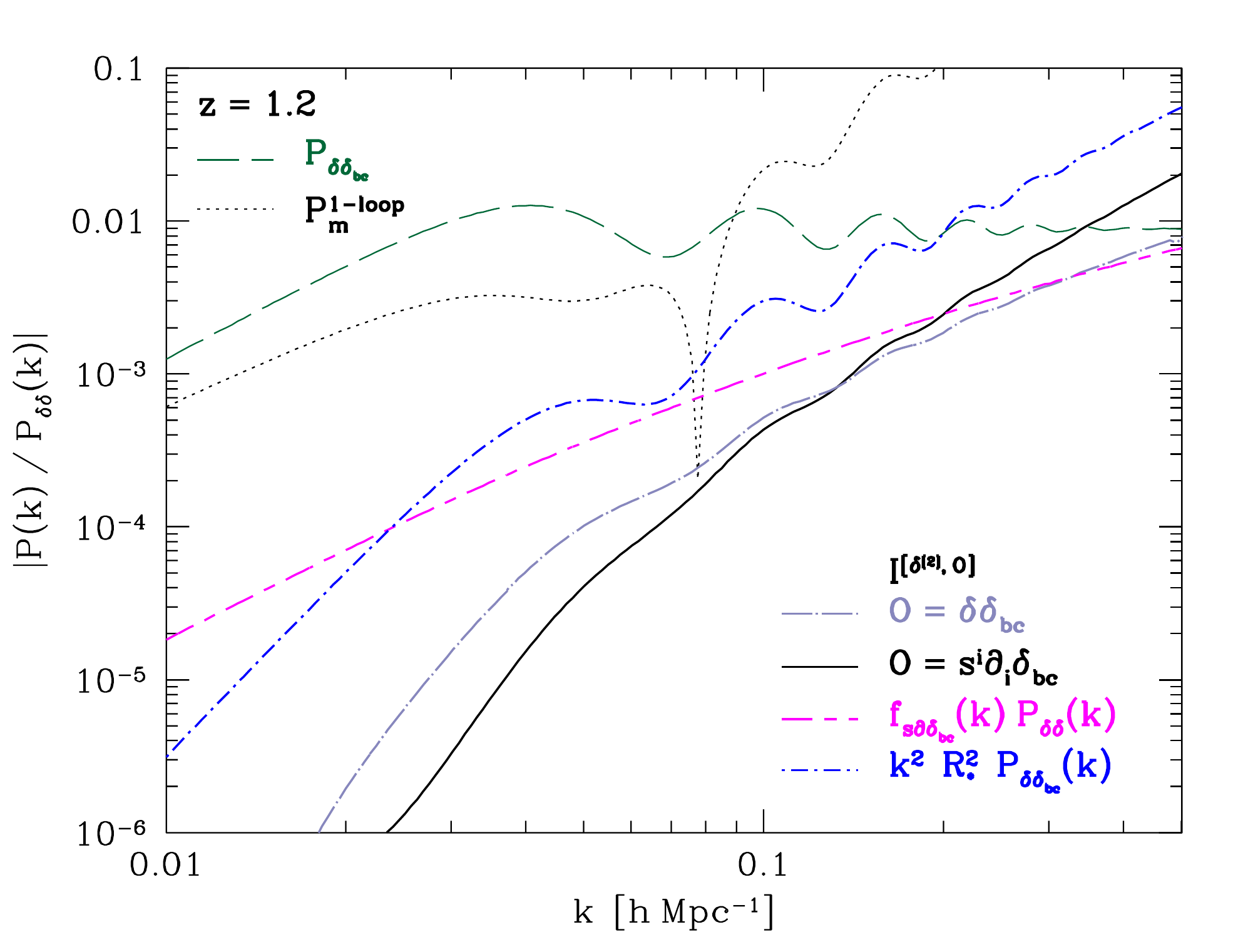}
\caption{Baryon-CDM relative velocity ($\propto \theta_{bc},\,v_{bc}^2$, left panel) and density ($\d_{bc}$, right panel; note the different scale) contributions to the galaxy power spectrum at 1-loop order (absolute magnitude at $z=1.2$), relative to the linear matter power spectrum.  Terms involving $\v{v}_{bc}^2$ have been scaled by $b^{bc}_{v^2} = 0.01 \s_{v_{bc}}^{-2}(z)$ [\refeq{bv2}], while terms involving $\theta_{bc}, \partial^i v_{bc}^j$ are scaled with $b_{\theta}^{bc} = 6.8/[(1+z) H_0]$ [\refeq{btheta}].  Terms involving $\d_{bc}$ in the right panel are scaled with $b_\d^{bc}=1$.  For $\d_{bc}$, we also show the leading higher derivative term (magenta short-long-dashed, assuming $R_* = 5\Mpch$).  For comparison, we also show the corresponding linear order contributions from \reffig{Pklin}, $P_{\d\theta_{bc}}(k)$ (left) and $P_{\d\d_{bc}}(k)$ (right), respectively, and illustrate the level of standard nonlinear contributions via the matter 1-loop power spectrum (dotted line). 
\label{fig:Pk1l}}
\end{figure*}
To summarize, at 1-loop order the effects of baryon-CDM perturbations on the galaxy power spectrum are fully generally described by five additional bias parameters: 
\be
\left\{b_{v^2}^{bc},\,b_{\d\d}^{bc},\,b_{\d\theta}^{bc},\,b_{\G}^{bc},\,b_{\rm nloc}^{bc}\right\}\,.
\ee
If we were to include terms of order $\d_{bc}^2,\,\theta_{bc}^2$ and higher, this would add another four bias parameters.  
Extending the notation introduced by \cite{assassi/etal}, we can succinctly
summarize the contributions to $P_{gg}$ as 
\ba
& P_{gg}^{\text{1-loop}}(k)\Big|_{bc} = 2 b_1 P_{gm}^{\text{1-loop}}(k)\Big|_{bc}\!\!
+ \sum_{O,O'_{bc}}\!\! b_O b_{O'_{bc}} \mathcal{I}^{[O,O'_{bc}]}(k)\,,
\nonumber
\ea
where $O \in \{\d^2,\, \G_2\}$, while
\be
O'_{bc} \in \{ v_{bc}^2,\,\d\d_{bc},\, \d\theta_{bc},\, \G^{bc},\,  
s^i \partial_i \delta_{bc},\,  s^i \partial_i \theta_{bc} \}\,,
\ee
with $b_{s\partial\delta_{bc}}\equiv -b_\delta^{bc}$,
$b_{s\partial\theta_{bc}}\equiv -b_\theta^{bc}$.  Finally, $P_{gm}$ is given by
\ba
 P_{gm}^{\text{1-loop}}(k)\Big|_{bc} =\:& b^{bc}_{\d \delta} \mathcal{I}^{[\d^{[2]}, \d\d_{bc}]}(k) 
+ b^{bc}_{\d \theta} \mathcal{I}^{[\d^{[2]}, \d\theta_{bc}]}(k) 
\vs
& - b^{bc}_{\delta} \mathcal{I}^{[\d^{[2]}, s\partial\delta_{bc}]}(k)
- b^{bc}_{\theta} \mathcal{I}^{[\d^{[2]}, s\partial\theta_{bc}]}(k) \vs
& + b^{bc}_{\G} \mathcal{I}^{[\d^{[2]}, \G^{bc}]}(k) + b^{bc}_{v^2} \mathcal{I}^{[\d^{[2]}, v_{bc}^2]}(k) \vs
& + \bigg[\left(b^{bc}_{\G} + \frac25 b^{bc}_{\rm nloc} \right)  f_{\theta_{bc}}(k) 
\label{eq:Pgm1l}\\
& \qquad- b^{bc}_{\theta} f_{s\partial\theta_{bc}}(k) 
- b^{bc}_{\delta} f_{s\partial\delta_{bc}}(k) \bigg] P_{\d\d}(k)  \vs
& + b_{\partial^2\d}^{bc} k^2 P_{\d\d_{bc}}(k) + b_{\partial^2\theta}^{bc} k^2 P_{\d\theta_{bc}}(k) 
\,.\nonumber
\ea
Here we have defined 
\ba
f_{Y}(k) =\:& 4 \int_{\vp} 
S_{\G_2}(\vk-\vp, \vp)
F_2(\vk,-\vp) P_{\d Y}(p)
\vs
f_{s\partial Y}(k) =\:& \frac37 \int_{\vp} 
\left[S_{s\partial\theta_{bc}}(\vk-\vp,\vp) - 1 \right] \vs
& \qquad\times S_{\G_2}(\vk,-\vp) P_{\d Y}(p)  \,,
\nonumber
\ea
where $Y = \d_{bc},\,\theta_{bc}$ and the kernels are given below.  Further,
\ba
\mathcal{I}^{[O, O'_{bc}]}(k) =\:& 2 \int_{\vp} \bigg[ S_O(\vp,\vk-\vp) S_{O'_{bc}}(\vp,\vk-\vp) \label{eq:defs} \\
& \qquad\  \times P_{\d X}(p) P_{\d Y}(|\vk-\vp|)\vs
& - S_O(\vp,-\vp) S_{O'_{bc}}(\vp,-\vp)  P_{\d X}(p) P_{\d Y}(p) \bigg] \,,
\nonumber
\ea
where $X = \theta_{bc}$ for $O'_{bc}=v_{bc}^2$, and $X=\d$ otherwise, while $Y = \theta_{bc}$ for operators involving $v_{bc}^i$, $\partial^i v_{bc}^j$ and $Y = \d_{bc}$ for operators involving $\d_{bc}$.  $F_2$ denotes the symmetrized perturbation theory kernel \cite{bernardeau/etal:2002}.  Finally, letting $\mu = \vk_1\cdot\vk_2/(k_1k_2)$,
\be
S_O(\vk_1,\vk_2) = \left\{ \begin{array}{cl}
1, & O = \d^2,\,\d\d_{bc},\,\d\theta_{bc} \\[1pt]
\mu^2 - 1, & O = \G_2,\,\G^{bc} \\[2pt]
F_2(\vk_1,\vk_2), & O = \d^{[2]} \\
-\mu\,k_2/k_1, & O = s^i \partial_i \delta_{bc},\, s^i \partial_i \theta_{bc} \\[1pt]
-\mu / (k_1 k_2), & O = v_{bc}^2
\end{array}
\right. .
\nonumber
\ee
In the definition of $\mathcal{I}^{[O,O']}$ and $f_{s\partial Y}$, we subtract the constant contribution for $k\to 0$ that is present for $O\neq \delta^{[2]}$, since it renormalizes the galaxy shot noise.  Note this is not done in \cite{yoo/seljak}, resulting in relatively large contributions from $v_{cb}^2$ at low $k$ which in practice are absorbed in the shot noise.    
We have neglected any contribution from baryon-CDM perturbations to the matter power spectrum $P_{mm}(k)|_{bc}$.  At low redshifts, these are expected to be much smaller than those introduced by galaxy biasing \cite{lewandowski/perko/senatore}.

\reffig{Pk1l} shows a subset of the terms in $P_{gg}^{\text{1-loop}}$.   We only show $\mathcal{I}^{[\d^{[2]},O'_{bc}]}$ since these
are slightly larger than the others, $\mathcal{I}^{[\d^2,O'_{bc}]},\,\mathcal{I}^{[\G_2,O'_{bc}]}$, while the scale dependence
is essentially identical.  The left panel shows terms involving $v_{bc}^2$,
$\partial^i v_{bc}^j$, while the right panel shows those containing $\d_{bc}$.  
In the latter case, we also show the leading higher derivative term
[last line in \refeq{Pgm1l}].   
Further, we do not show all terms for $\d_{bc}$, since their scale dependence is very similar to that of the corresponding term involving $\theta_{bc}$.  Each term is to be multiplied by one of $\{b_{\theta}^{bc},\,b_{v^2}^{bc},\,b_{\d\theta}^{bc},\,b_{\G}^{bc},\,b_{\rm nloc}^{bc}\}$, for which we do not have precise individual estimates.  Instead, we rely on the scaling described at the end of \refsec{biasgen}, multiplying each term by either $b_\d^{bc}=1$ (operators $\propto \d_{bc}$), $b^{bc}_{v^2}$ (operators $\propto v_{bc}^2$) or $b_\theta^{bc}$ (operators $\propto \theta_{bc},\,\partial_i v_{bc}^j$).  
The 1-loop contributions are smaller than
the linear order contributions on large scales, while the 1-loop terms involving
$\v{v}_{bc}^2$ and $\theta_{bc}$ are comparable.  This validates the counting of baryon-CDM contributions, and confirms that we can neglect the higher order terms in $\d_{bc},\,\theta_{bc}$.  However, note that this ranking assumes   $b_{v^2}^{bc}\sim 0.01 \s_{v_{bc}}^{-2}$ [\refeq{bv2}]; 
a significantly smaller value of $b_{v^2}^{bc}$ will make the terms 
$\propto \theta_{bc},\,\partial_i v_{bc}^j$ relatively more important.

\section{Galaxy velocities}
\label{sec:vel}

In order to describe the observed statistics of galaxies, we also need
a description of galaxy velocities, since they affect the statistics via
redshift-space distortions (see \cite{hamilton:1998} for a review).  Thus, we now derive the contributions
of baryon-CDM perturbations to galaxy velocities.  
The relation between the velocities of galaxies and that of matter, referred to as \emph{velocity bias}, is severely constrained by the equivalence principle.   
Specifically, in the bias expansion of the local relative velocity between galaxies and matter,
\be
\vv_{g,\rm rel}(\vx,\tau) = \vv_g(\vx,\tau) - \vv_m(\vx,\tau)\,,
\ee
we can only allow for local observables to appear, just as in the expansion of the galaxy density field.  Since $\vv_{g,\rm rel}$ is a vector, this implies
that only higher-derivative terms can contribute for adiabatic growing-mode initial conditions, specifically $\vv_{g,\rm rel} \propto \partial\partial\partial\Phi$ \cite{senatore:2014,MSZ}.  

At linear order in perturbations, there are three contributions from baryon-CDM perturbations up to second order in derivatives, namely
\ba
\vv_{g,\rm rel}^{(1)}(\vx,\tau) =\:& \beta_v^{bc}(\tau) \vv_{bc}(\vq) \vs
& + \beta_{\partial\d}^{bc}(\tau) \boldsymbol{\nabla} \d_{bc}(\vq)
+ \beta_{\partial\theta}^{bc}(\tau) \boldsymbol{\nabla} \theta_{bc}(\vq)
\,,
\label{eq:vrellin}
\ea
where we denote coefficients in the bias expansion of $\vv_{g,\rm rel}$ with
$\beta_O$.  Here, one expects the second and third terms to scale as $R_*^2$, 
where $R_*$ is the spatial nonlocality scale of galaxies.  On the other hand,
$\beta_v^{bc}$ does not necessarily scale with $R_*$ but is expected to be at most of order one.  To see this, 
recall that at linear order, the baryon and CDM velocities are each given
by a linear combination of $\vv_{bc}$ and $\vv_m$ [see above \refeq{DDtau}].  
Thus, if $\beta_v^{bc} = - f_b$, then galaxies comove with the baryon fluid
$\vv_g = \vv_b$.  Conversely, if $\beta_v^{bc} = 1-f_b$, then $\vv_g = \vv_c$ corresponding to galaxies comoving with CDM.  One would expect these two
cases to bracket the physically reasonable range of $\beta_v^{bc}$.  

This term is unique to the baryon-CDM relative velocity mode.  
Note that there is no contradiction with the statements regarding velocity
bias made above:  first, the baryon-CDM perturbations are imprinted
in the initial conditions, whereas the statements of \cite{senatore:2014,MSZ}
apply to velocity bias induced by nonlinear evolution from adiabatic
initial conditions.  Second, baryon-CDM perturbations are themselves
induced by causal physics around decoupling.  For scales
much larger than the sound horizon at decoupling (and neglecting any
contributions from reionization), $\vv_{bc}$ is indeed
proportional to three derivatives of the potential in agreement with the
argument made above, as can easily be verified using the output of
Boltzmann codes.

Let us consider the contribution of the first term in \refeq{vrellin} to
the linear galaxy power spectrum in redshift space:
\ba
P_{gg}^{s,\rm lin}(k,\mu) = (b_1 + f\mu^2) \Big[& (b_1 + f\mu^2) P_{\d\d}(k)
+ 2 b_\theta^{bc} P_{\d\theta_{bc}}(k) \vs
& + b_\d^{bc} P_{\d\d_{bc}}(k) - \frac{\beta_v^{bc}}{aH} \mu^2 P_{\d\theta_{bc}}(k) \Big]\,, \nonumber
\ea
where $f \equiv d\ln D/d\ln a$ is the linear growth rate, and $\mu$ is the
angle of $\vk$ with the line of sight.  
Here we have again neglected terms quadratic in baryon-CDM perturbations.  
The term $\propto P_{\d\d}(k)$ is the standard Kaiser redshift-space
power spectrum \cite{kaiser:1987}.  Clearly, the baryon-CDM contributions
lead to a bias in the estimate of the growth rate $f$ using the large-scale galaxy
power spectrum if not accounted for.  

The second and third term in \refeq{vrellin} are analogous to the higher derivative terms encountered
in the bias expansion for adiabatic modes.  Note that $\partial^i \theta_{bc} = \partial^2 v_{bc}^i$ since $\vv_{bc}$ is longitudinal.  
Physically, they are induced by the nonlocality of galaxy formation discussed in \refsec{bthetabv2}; both
$\beta_{\partial\d}^{bc}$ and $\beta_{\partial\theta}^{bc}$ scale as $R_*^2$,
as any filtering on the scale $R_*$ in the absence of preferred directions
has to scale quadratically with the filter scale.  Thus, we expect
$\beta_{\partial\theta}^{bc} \sim R_*^2$, while $\beta_{\partial\delta}^{bc} \sim \cH R_*^2$, where the latter estimate is based on adding a term $b_{\partial^2\d}^{bc} \partial^2 \d_{bc}$ to $\d_g$ and then using the approximate scaling based 
on a continuity equation for galaxies, $\theta_g = -\d_g'$.  

At nonlinear order, we have to allow for all combinations of local 
observables that transform like a vector to appear in the expansion of
$\vv_{g,\rm rel}$.  That is, these terms should be constructed
out of $R_+,\,\v{R}_-,\,\partial^i R_-^j$ and $\Pi_{ij}^{[n]}$ introduced
in \refsec{biasgen} and spatial derivatives thereof.  Restricting to terms second order in perturbations,
but linear in baryon-CDM contributions, and at the same order in derivatives
as \refeq{vrellin}, we have
\ba
v_{g,\rm rel}^i \Big|_{\rm 2nd} \supset\:& v_{bc}^i {\rm Tr}[\Pi^{[1]}]\,,\  
v_{bc}^j \Pi^{[1]}_{ij}\,,\  \partial^i\d_{bc} {\rm Tr}[\Pi^{[1]}]\,,\vs
& \partial^j \d_{bc} \Pi^{[1]}_{ij}\,,\   
\partial^i \theta_{bc} {\rm Tr}[\Pi^{[1]}]\,,\  
\partial^j \theta_{bc} \Pi^{[1]}_{ij}\,,\  \vs
&\partial^i \partial^k v_{bc}^m \Pi^{[1]}_{km}\,,\  
\partial^2 v_{bc}^m \Pi^{[1]}_{im}\,,
\ea
where as always $\Pi^{[1]}$ is evaluated at $(\vx,\tau)$ while $\vv_{bc},\,\d_{bc}$ are evaluated at $\vq$.  Here, the first two terms are expected to have bias coefficients of order one, while the others scale as $R_*^2$.  Clearly,
the number of higher derivative terms multiply rapidly at higher order,
as is the case in the expansion for adiabatic modes only \cite{mcdonald/roy:2009}.

\begin{figure}[t!]
\centering
\includegraphics[width=0.49\textwidth]{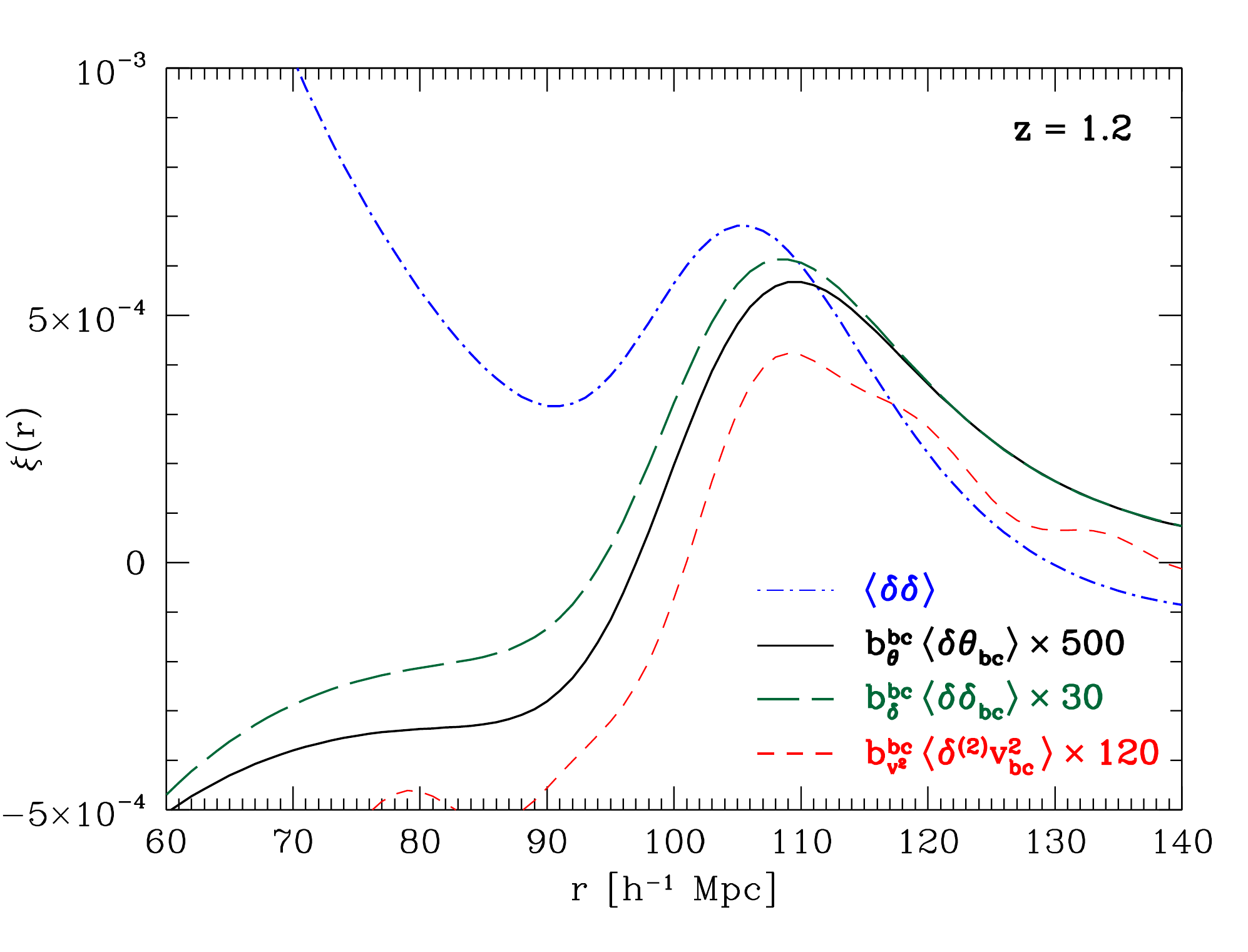}
\vspace*{-0.5cm}
\caption{The three leading contributions $\propto \d_{bc},\,\theta_{bc},$
and $v_{bc}^2$ to the galaxy correlation function in real space, i.e. the real-space version of the results shown in \reffig{Pklin} (same line styles).  We have scaled each contribution by the number indicated, in order to illustrate the shape and position of the BAO feature in each contribution.  The term $b_\d^{bc} P_{\d\d_{bc}}(k)$ is expected to yield the most significant systematic shift of the BAO scale.
\label{fig:xi}}
\end{figure}

\section{Conclusions}
\label{sec:concl}

An accurate, robust theoretical model for the statistics of galaxy
counts (two- and higher-point functions) on large scales is a crucial
ingredient in order to extract cosmological constraints from large
redshift surveys.  Beyond the frequently considered adiabatic growing
mode of the total matter (baryon+CDM) fluid, relative density and velocity
perturbations between baryons and CDM, which are induced by pre-recombination
plasma oscillations, also need to be taken into account as they lead to
percent-level corrections to galaxy clustering on scales of tens to hundreds of Mpc.

The relative velocity perturbation was first pointed out by
\cite{tseliakhovich/hirata:2010}, while the relative density perturbation
was shown to exist in \cite{barkana/loeb:11}.  The contributions of
the relative-velocity perturbation to galaxy statistics
on large scales were studied in \cite{dalal/etal:2010,yoo/dalal/seljak,yoo/seljak,slepian/eisenstein,blazek/etal:15}.   
Refs.~\cite{dalal/etal:2010,yoo/dalal/seljak,yoo/seljak,slepian/eisenstein} only considered the term $\d_g \supset b_{v^2}^{bc} v_{bc}^2$.
Recently, this was applied in \cite{slepian/etal:2016} to the galaxy three-point function of the BOSS CMASS sample to put an upper limit on $b_{v^2}^{bc}$.  
However, Ref.~\cite{blazek/etal:15} pointed out that there are additional terms induced by the fact that the relative velocity amplitude has to be evaluated at the Lagrangian position.  In particular, they effectively obtained a linear order term $b_\theta^{bc} \theta_{bc}$ which had previously not been considered.

Here, building on recent results on the general bias expansion for
galaxies and other tracers \cite{senatore:2014,MSZ}, 
we have given a complete and consistent perturbative description of the effect of baryon-CDM perturbations on galaxy clustering, which contain the effects introduced by \cite{tseliakhovich/hirata:2010} and \cite{blazek/etal:15} as well as \cite{barkana/loeb:11}.  We perform a consistent renormalization of the bias parameters (as first proposed by \cite{mcdonald:2006}), and show that the term $b_\theta^{bc} \theta_{bc}$, treated as third order in \cite{blazek/etal:15}, is to be considered a \emph{linear order} contribution to the galaxy density.  Note that
this implies that there are additional contributions to the leading galaxy three-point function as well, in particular $\d\,\theta_{bc}$ and $s^i \partial_i \theta_{bc}$ and the corresponding terms involving $\d_{bc}$, which have so far not been included in \cite{slepian/eisenstein,slepian/etal:2016}.

We also explicitly derive, for the first time, the baryon-CDM contributions to the galaxy velocity (\refsec{vel}), which are important to model observed galaxy statistics which are affected by redshift-space distortions.

We then use the general bias expansion to derive the complete contributions to the linear (\refsec{lin}) and 1-loop (\refsec{1loop}) galaxy two-point functions.  Three physical effects can be distinguished: the constant compensated mode $\d_{bc}$, which effectively modulates the local baryon-CDM ratio;  divergence and shear in the relative velocity $\theta_{bc},\,\partial^i v_{bc}^j$, where both $\d_{bc}$ and $\theta_{bc}$ enter at linear order;  and the relative velocity itself, which enters as $v_{bc}^2$ at lowest order (in the galaxy density;  it does enter at linear order in the galaxy \emph{velocity}).   
Using physical models of the bias parameters (\refsec{biaslin}) for the $\d_{bc}$ and $\theta_{bc}$ contributions, we estimate that the constant compensated mode $\d_{bc}$ is the most important baryon-CDM effect on galaxy clustering by a significant margin on large scales.  The bias $b_{v^2}^{bc}$ is the most theoretically uncertain, and this contribution could be either larger or smaller than the velocity divergence effect which is estimated as $b_\theta^{bc} \sim 7 H_0^{-1} (b_1-1)/(1+z)$.  
The terms appearing at 1-loop order are generally even smaller than the linear order contributions.  However, it is important to include all, if any, of the 1-loop contributions in order to have a consistent model.  \refsec{1loop} provides all the necessary expressions for this (although redshift-space distortions also need to be taken into account).  

Of particular importance are the baryon acoustic oscillation (BAO) features imprinted on the baryon-CDM contributions, which are able to shift the BAO feature in the galaxy power spectrum, and thus systematically bias this standard ruler.   
Note that this applies in principle to all three contributions $\propto \d_{bc},\,\theta_{bc},\, v_{bc}^2$ shown in Fourier space in \reffig{Pklin}.  This is illustrated in \reffig{xi}, which shows these terms in real space.  Clearly, all three contributions exhibit a prominent BAO feature that is shifted, to varying degrees, from that in the growing mode.  Nevertheless, none of these contributions is expected to exceed the 1\% level, so that current BAO constraints are very likely to be unbiased within error bars.
  Marginalizing over the leading additional bias parameters introduced here will remove any potential systematic bias introduced by primordial baryon-CDM perturbations in future, larger surveys.   
Beyond the BAO feature, the baryon-CDM perturbations also change the
broad-band shape of the galaxy power spectrum, and change the anisotropy
of the power spectrum in redshift space, as discussed in \refsec{vel}.  Thus,
taking these contributions into account is just as important for measurements
of the growth rate $f = d\ln D/d\ln a$ from galaxy redshift surveys.

For the practical analysis, physical estimates of the bias parameters such as those presented in \refsec{biaslin} and \refsec{vel} will be highly useful.  Clearly, there is significant room for refining these estimates through both semi-analytical and fully numerical approaches.  

\acknowledgments
I would like to thank Kyungjin Ahn, Florian Beutler, Jonathan Blazek, Pat~McDonald, Mehrdad~Mirbabayi, Shun~Saito, Uro$\check{\rm s}$ Seljak, Jaiyul Yoo, Drian van der Woude, and Simon White for helpful discussions, and acknowledge support from the Marie Curie Career Integration Grant  (FP7-PEOPLE-2013-CIG) ``FundPhysicsAndLSS.''

\appendix

\section{Transfer functions}
\label{app:Tk}

The linear predictions for the total matter density perturbation as well 
as the baryon-CDM perturbations can be taken from the Boltzmann solver
CAMB \cite{camb}.  However, some subtleties need to be observed when taking 
CAMB transfer functions as initial
conditions for perturbation theory calculations or N-body simulations.  
The latter calculations assume matter (or dark energy) domination, and ignore
radiation as well as neutrinos, in particular the significant neutrino
anisotropic stress at early times.  Suppose one took the CAMB transfer
function at some high redshift $z_{\rm in} \gtrsim 50$.  Then, standard
perturbation theory and N-body codes would not yield the correct \emph{linear}
evolution of perturbations.  Moreover, results will depend on the 
initial redshift chosen (see \cite{somogyi/smith:2010} for a nice description of this issue).  Thus, in order to ensure that the 
correct linear evolution at low redshifts is recovered, the following
procedure should be adopted:
\begin{enumerate}
\item Obtain CAMB transfer functions $T_b(k)$, $T_c(k)$ for baryons and CDM, as well as $T_{v_{bc}}(k)$ for $v_{bc}$, at low redshifts.  Here we assume $z=0$.
\item Match to the linear modes derived in \refsec{intro},
\ba
R_-(k) =\:& \frac{k}{H_0} \frac{T_{v_{bc}}(k)}{T_m(k)}\, \d_m(k,0) \vs
R_+(k) =\:& \frac{T_b(k)-T_c(k)}{T_m(k)} \d_m(k,0) - R_-(k) D_r(\tau_0)\,,
\nonumber
\ea
where $\d_m(k,z=0) \propto T_m(k) A_s^{1/2} $, $T_m = f_b T_b + (1-f_b) T_c$, is the standard growing-mode total matter density perturbation.  In practice, the second term $\propto R_-$ in $R_+$ is a small (less than 1\%) correction.
\item Using the linear solution that is exact for matter+$\Lambda$ [\refeq{drgen}], calculate $\d_m(k,z_{\rm in}), \d_r(k,z_{\rm in})$ at the desired initial redshift $z_{\rm in}$.  
\end{enumerate}
$\d_b(k,z_{\rm in}),\d_c(k,z_{\rm in})$ are then trivial linear combinations of $\d_m, \d_r$ and can be used as initial conditions for N-body codes or perturbative calculations.   Only this particular matching is guaranteed to recover the correct linear evolution at low redshifts.

\section{Spherical collapse}
\label{app:scollapse}

In this appendix we provide more details on the spherical collapse calculation of \refsec{scollapse}.  
Following the notation of App.~A in \cite{HPMhalopaper}, we replace $t$
with $\ln a$ as time coordinate and introduce
\ba
y_s(a) =\:& \frac{R_s(a)}{R_{\rm in}} - \frac{a}{a_{\rm in}} \,,\quad s=c,b\,,
\ea
where the second term subtracts out the Hubble flow.  This implies that $y_s(a_{\rm in}) = 0$ and
\be
1 + \d_s(a) = [1 + \d_s(a_{\rm in}) ] \left[ \frac{a_{\rm in}}{a} y_s(a) + 1\right]^{-3}\,.
\label{eq:deltaofy}
\ee
We choose $a_{\rm in} = a_{\rm dec} = 10^{-3}$.  Further, we have
\be
\frac{R_b}{R_c} = \frac{y_b + a/a_{\rm in}}{y_c + a/a_{\rm in}}\,.
\label{eq:RboverRc}
\ee
Denoting derivatives with respect to $\ln a$ as primes, 
the equation for $y_s$ then is (this is a generalization of Eq.~(A9) in \cite{HPMhalopaper})
\ba
& y_s'' + \frac{H'}{H} y_s' -\left(1 + \frac{H'}{H}\right) y_s =  - \frac12 \Om(a) \left(\frac{a}{a_{\rm in}}+y_s\right) G_s \vs
& G_s = \sum_{t=c,b} f_t 
\left\{
\begin{array}{cc}
\d_t \,, & R_t \leq R_s \\[4pt]
{\rm max}\{0, (1+\d_t) (R_s/R_t)^3 - 1 \}\,, & R_s < R_t \\
\end{array} \right. \,.
\label{eq:ybc}
\ea
The source term $G_s$ is continuous, but not
differentiable at $R_b=R_c$, where it reduces to $G_b = G_c = \d_m$.  
This means that the response of the tophat evolution to $\theta_{bc}$ around $\theta_{bc}=0$ is ill-defined, since
we attempt to take a derivative at precisely this point.  For this reason,
we perform a continuously differentiable cubic spline interpolation between
the two branches in the interval $1 \leq R_b/R_c \leq 1.1$ (we have verified
that the upper end of the range has negligible impact on the results).  This corresponds
to slightly smoothing the tophat density profiles of the shells.  Crucially,
the interpolation matches the exact result at $R_b=R_c$, and so ensures that
the evolution is correct at linear order.

We now derive the initial conditions.  At $a_{\rm in}$ we wish to impose
\ba
\d_c(a_{\rm in}) =\:& \d_{c,\rm dec}; \quad \d_b(a_{\rm in})\Big|_{\theta_{bc}=0} = f_{\d,\rm dec} \d_{c,\rm dec}\,,
\ea
where we use linear theory for the initial conditions, and $\d_c$ is assumed to be in the growing mode.  Our default choice is $f_{\d,\rm dec}=0$ (\refsec{scollapse}), but we allow for nonzero values here.  
Using that
\ba
\d_{c,\rm dec} =\:& A_+ D_+(\tau_{\rm dec}) 
\label{eq:dadec}\\
\d_{b,\rm dec} - \d_{c,\rm dec} =\:& R_+ + \frac{\theta_{bc,0}}{H_0} D_r(\tau_{\rm dec})
\nonumber
\ea
we can then trade the three mode amplitudes $A_+,\,R_+,\,R_-$ for 
$\d_{c,\rm dec},\,f_{\d,\rm dec},\,\theta_{cb,0}$ via
\be
R_+ = \frac{f_{\d,\rm dec}-1}{1+(f_{\d,\rm dec}-1) f_b} \d_{c,\rm dec}\,.
\label{eq:Rplusin}
\ee
Note that even for $f_{\d,\rm dec}=1$ ($R_+=0$), $\d_{b,\rm dec}\neq \d_{c,\rm dec}$ due to the second term in \refeq{dadec}.  Neglecting this term would lead to a spurious constant contribution to $\d_b-\d_c$ which is proportional to $(\theta_{bc}/\cH)_{\rm dec}$.  
Finally, note that
$\theta_s = 3\cH (H^{-1} \dot R_s/R_s - 1)$.  We thus have
$y'_s(a_{\rm in}) = -\d_s'(a_{\rm in})/3$.  With these relations the initial conditions
for the spherical collapse are completely specified in terms of
$\d_{c,\rm dec},\,f_{\d,\rm rec},\,\theta_{cb,0}$.  
\begin{figure}[t!]
\centering
\includegraphics[width=0.49\textwidth]{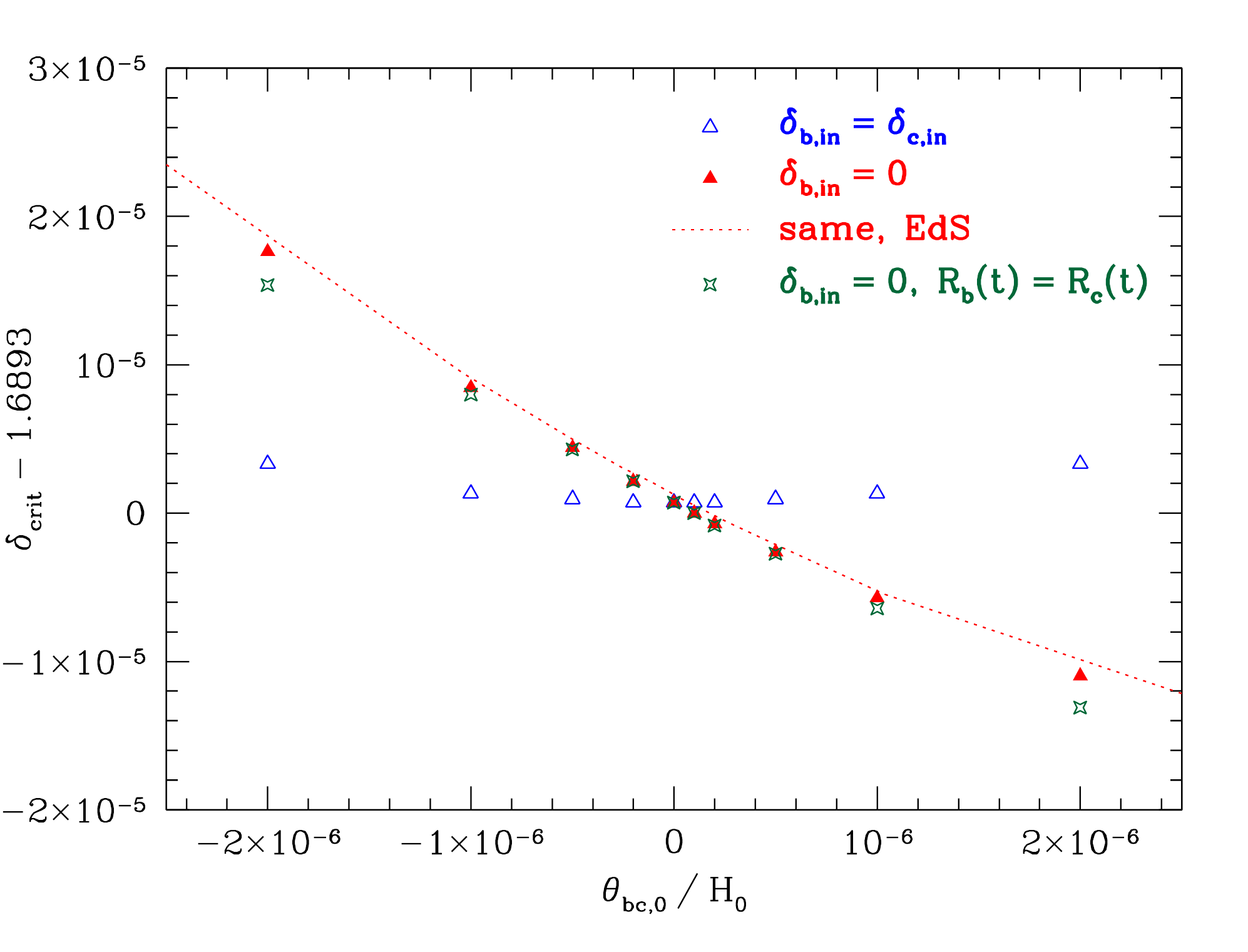}
\vspace*{-0.5cm}
\caption{Collapse threshold $\d_{\rm crit}(z=1.2)$ as a function of $\theta_{bc,0}/H_0$ for different initial conditions and cosmologies (see text).  The red filled triangles correspond to the default case discussed in \refsec{scollapse}.
\label{fig:dcrit}}
\end{figure}

\reffig{dcrit} shows the resulting collapse threshold as a function of
$\theta_{bc,0}/H_0$.  The red filled triangles show the default case
($f_{\d,\rm dec}=0$),
while the dotted line shows the result for an Einstein-de Sitter universe.  
Specifically, we set $\Om \to 1$ while adjusting $\Omega_b$ to keep 
$f_b$ at the same value as in the fiducial cosmology.  Further, since
the physically relevant quantity is $(\theta_{bc}/\cH)_{\rm dec}$, we
plot the Einstein-de Sitter result as a function of $\sqrt{\Om} \theta_{bc,0}/H_0$, where $\Om=0.27$ is the fiducial value.  We see that the effect
of $\theta_{bc}$ on $\d_{\rm crit}$ is only very weakly dependent on 
cosmology.  The green stars in \reffig{dcrit} show the result obtained
when forcing the same gravity to act on both shells, by setting
$R_b=R_c$ in the evaluation of $G_s$ in \refeq{ybc}.  This shows that
our results are insensitive to the details of the spline interpolation used in the
evaluation of \refeq{ybc}.  Finally, the
blue open triangles in \reffig{dcrit} show the result for setting
$f_{\d,\rm dec}=1$.  In this case, both matter components have the same
overdensity initially and only differ via $\dot R_b \neq \dot R_c$
(recall that both shells always have the same initial radius).  
In this case, $\theta_{bc}=0$ corresponds approximately to a local
minimum in $\d_{\rm crit}$.  As argued in \refsec{scollapse} however,
we expect $f_{\d,\rm dec}=0$ to be the physically relevant case.
 
In the previous preprint version of this paper, we reported a slope
$\partial\d_{\rm crit}/\partial(\theta_{bc,0}/H_0)$ that is roughly
one order of magnitude higher than the value given here.  That result
is incorrect and affected by three issues.  
First, the previous implementation of initial conditions
included a spurious contribution to $R_+$ proportional to $(\theta_{bc}/\cH)_{\rm dec}$ [see discussion after \refeq{Rplusin}].  Second, different initial radii
of the baryon and CDM shells were used, which results in an evolution
that does not follow linear theory initially.  Finally, the previous 
calculation used the non-differentiable force term in \refeq{ybc}, 
i.e. without spline interpolation, which further increases the slope (the
different initial radii shift the discontinuity in the derivative
$\partial\d_{\rm crit}/\partial(\theta_{bc,0}/H_0)$ away from $\theta_{bc,0}=0$).

\clearpage
\bibliography{REFS}

\begin{thebibliography}{42}
\expandafter\ifx\csname natexlab\endcsname\relax\def\natexlab#1{#1}\fi
\expandafter\ifx\csname bibnamefont\endcsname\relax
  \def\bibnamefont#1{#1}\fi
\expandafter\ifx\csname bibfnamefont\endcsname\relax
  \def\bibfnamefont#1{#1}\fi
\expandafter\ifx\csname citenamefont\endcsname\relax
  \def\citenamefont#1{#1}\fi
\expandafter\ifx\csname url\endcsname\relax
  \def\url#1{\texttt{#1}}\fi
\expandafter\ifx\csname urlprefix\endcsname\relax\def\urlprefix{URL }\fi
\providecommand{\bibinfo}[2]{#2}
\providecommand{\eprint}[2][]{\url{#2}}

\bibitem[{\citenamefont{{Eisenstein} et~al.}(2005)\citenamefont{{Eisenstein},
  {Zehavi}, {Hogg}, {Scoccimarro}, {Blanton}, {Nichol}, {Scranton}, {Seo},
  {Tegmark}, {Zheng} et~al.}}]{BAO/SDSS}
\bibinfo{author}{\bibfnamefont{D.~J.} \bibnamefont{{Eisenstein}}},
  \bibinfo{author}{\bibfnamefont{I.}~\bibnamefont{{Zehavi}}},
  \bibinfo{author}{\bibfnamefont{D.~W.} \bibnamefont{{Hogg}}},
  \bibinfo{author}{\bibfnamefont{R.}~\bibnamefont{{Scoccimarro}}},
  \bibinfo{author}{\bibfnamefont{M.~R.} \bibnamefont{{Blanton}}},
  \bibinfo{author}{\bibfnamefont{R.~C.} \bibnamefont{{Nichol}}},
  \bibinfo{author}{\bibfnamefont{R.}~\bibnamefont{{Scranton}}},
  \bibinfo{author}{\bibfnamefont{H.-J.} \bibnamefont{{Seo}}},
  \bibinfo{author}{\bibfnamefont{M.}~\bibnamefont{{Tegmark}}},
  \bibinfo{author}{\bibfnamefont{Z.}~\bibnamefont{{Zheng}}},
  \bibnamefont{et~al.}, \bibinfo{journal}{\apj} \textbf{\bibinfo{volume}{633}},
  \bibinfo{pages}{560} (\bibinfo{year}{2005}), \eprint{astro-ph/0501171}.

\bibitem[{\citenamefont{{Cole} et~al.}(2005)\citenamefont{{Cole}, {Percival},
  {Peacock}, {Norberg}, {Baugh}, {Frenk}, {Baldry}, {Bland-Hawthorn},
  {Bridges}, {Cannon} et~al.}}]{BAO/2dF}
\bibinfo{author}{\bibfnamefont{S.}~\bibnamefont{{Cole}}},
  \bibinfo{author}{\bibfnamefont{W.~J.} \bibnamefont{{Percival}}},
  \bibinfo{author}{\bibfnamefont{J.~A.} \bibnamefont{{Peacock}}},
  \bibinfo{author}{\bibfnamefont{P.}~\bibnamefont{{Norberg}}},
  \bibinfo{author}{\bibfnamefont{C.~M.} \bibnamefont{{Baugh}}},
  \bibinfo{author}{\bibfnamefont{C.~S.} \bibnamefont{{Frenk}}},
  \bibinfo{author}{\bibfnamefont{I.}~\bibnamefont{{Baldry}}},
  \bibinfo{author}{\bibfnamefont{J.}~\bibnamefont{{Bland-Hawthorn}}},
  \bibinfo{author}{\bibfnamefont{T.}~\bibnamefont{{Bridges}}},
  \bibinfo{author}{\bibfnamefont{R.}~\bibnamefont{{Cannon}}},
  \bibnamefont{et~al.}, \bibinfo{journal}{\mnras}
  \textbf{\bibinfo{volume}{362}}, \bibinfo{pages}{505} (\bibinfo{year}{2005}),
  \eprint{astro-ph/0501174}.

\bibitem[{\citenamefont{{Bernardeau} et~al.}(2002)\citenamefont{{Bernardeau},
  {Colombi}, {Gazta{\~n}aga}, and {Scoccimarro}}}]{bernardeau/etal:2002}
\bibinfo{author}{\bibfnamefont{F.}~\bibnamefont{{Bernardeau}}},
  \bibinfo{author}{\bibfnamefont{S.}~\bibnamefont{{Colombi}}},
  \bibinfo{author}{\bibfnamefont{E.}~\bibnamefont{{Gazta{\~n}aga}}},
  \bibnamefont{and}
  \bibinfo{author}{\bibfnamefont{R.}~\bibnamefont{{Scoccimarro}}},
  \bibinfo{journal}{\physrep} \textbf{\bibinfo{volume}{367}},
  \bibinfo{pages}{1} (\bibinfo{year}{2002}), \eprint{arXiv:astro-ph/0112551}.

\bibitem[{\citenamefont{{Grin} et~al.}(2011)\citenamefont{{Grin}, {Dor{\'e}},
  and {Kamionkowski}}}]{grin/dore/kamionkowski}
\bibinfo{author}{\bibfnamefont{D.}~\bibnamefont{{Grin}}},
  \bibinfo{author}{\bibfnamefont{O.}~\bibnamefont{{Dor{\'e}}}},
  \bibnamefont{and}
  \bibinfo{author}{\bibfnamefont{M.}~\bibnamefont{{Kamionkowski}}},
  \bibinfo{journal}{\prd} \textbf{\bibinfo{volume}{84}}, \bibinfo{eid}{123003}
  (\bibinfo{year}{2011}), \eprint{1107.5047}.

\bibitem[{\citenamefont{{Barkana} and {Loeb}}(2011)}]{barkana/loeb:11}
\bibinfo{author}{\bibfnamefont{R.}~\bibnamefont{{Barkana}}} \bibnamefont{and}
  \bibinfo{author}{\bibfnamefont{A.}~\bibnamefont{{Loeb}}},
  \bibinfo{journal}{\mnras} \textbf{\bibinfo{volume}{415}},
  \bibinfo{pages}{3113} (\bibinfo{year}{2011}), \eprint{1009.1393}.

\bibitem[{\citenamefont{{Soumagnac} et~al.}(2016)\citenamefont{{Soumagnac},
  {Barkana}, {Sabiu}, {Loeb}, {Ross}, {Abdalla}, {Balan}, and
  {Lahav}}}]{soumagnac/etal:16}
\bibinfo{author}{\bibfnamefont{M.~T.} \bibnamefont{{Soumagnac}}},
  \bibinfo{author}{\bibfnamefont{R.}~\bibnamefont{{Barkana}}},
  \bibinfo{author}{\bibfnamefont{C.~G.} \bibnamefont{{Sabiu}}},
  \bibinfo{author}{\bibfnamefont{A.}~\bibnamefont{{Loeb}}},
  \bibinfo{author}{\bibfnamefont{A.~J.} \bibnamefont{{Ross}}},
  \bibinfo{author}{\bibfnamefont{F.~B.} \bibnamefont{{Abdalla}}},
  \bibinfo{author}{\bibfnamefont{S.~T.} \bibnamefont{{Balan}}},
  \bibnamefont{and} \bibinfo{author}{\bibfnamefont{O.}~\bibnamefont{{Lahav}}},
  \bibinfo{journal}{ArXiv e-prints}  (\bibinfo{year}{2016}),
  \eprint{1602.01839}.

\bibitem[{\citenamefont{{Lewandowski} et~al.}(2015)\citenamefont{{Lewandowski},
  {Perko}, and {Senatore}}}]{lewandowski/perko/senatore}
\bibinfo{author}{\bibfnamefont{M.}~\bibnamefont{{Lewandowski}}},
  \bibinfo{author}{\bibfnamefont{A.}~\bibnamefont{{Perko}}}, \bibnamefont{and}
  \bibinfo{author}{\bibfnamefont{L.}~\bibnamefont{{Senatore}}},
  \bibinfo{journal}{\jcap} \textbf{\bibinfo{volume}{5}}, \bibinfo{eid}{019}
  (\bibinfo{year}{2015}), \eprint{1412.5049}.

\bibitem[{\citenamefont{{Shoji} and {Komatsu}}(2009)}]{shoji/komatsu}
\bibinfo{author}{\bibfnamefont{M.}~\bibnamefont{{Shoji}}} \bibnamefont{and}
  \bibinfo{author}{\bibfnamefont{E.}~\bibnamefont{{Komatsu}}},
  \bibinfo{journal}{\apj} \textbf{\bibinfo{volume}{700}}, \bibinfo{pages}{705}
  (\bibinfo{year}{2009}), \eprint{0903.2669}.

\bibitem[{\citenamefont{{Somogyi} and {Smith}}(2010)}]{somogyi/smith:2010}
\bibinfo{author}{\bibfnamefont{G.}~\bibnamefont{{Somogyi}}} \bibnamefont{and}
  \bibinfo{author}{\bibfnamefont{R.~E.} \bibnamefont{{Smith}}},
  \bibinfo{journal}{\prd} \textbf{\bibinfo{volume}{81}}, \bibinfo{eid}{023524}
  (\bibinfo{year}{2010}), \eprint{0910.5220}.

\bibitem[{\citenamefont{{Bernardeau} et~al.}(2013)\citenamefont{{Bernardeau},
  {Van de Rijt}, and {Vernizzi}}}]{bernardeau/vdr/vernizzi}
\bibinfo{author}{\bibfnamefont{F.}~\bibnamefont{{Bernardeau}}},
  \bibinfo{author}{\bibfnamefont{N.}~\bibnamefont{{Van de Rijt}}},
  \bibnamefont{and}
  \bibinfo{author}{\bibfnamefont{F.}~\bibnamefont{{Vernizzi}}},
  \bibinfo{journal}{\prd} \textbf{\bibinfo{volume}{87}}, \bibinfo{eid}{043530}
  (\bibinfo{year}{2013}), \eprint{1209.3662}.

\bibitem[{\citenamefont{{Tseliakhovich} and
  {Hirata}}(2010)}]{tseliakhovich/hirata:2010}
\bibinfo{author}{\bibfnamefont{D.}~\bibnamefont{{Tseliakhovich}}}
  \bibnamefont{and} \bibinfo{author}{\bibfnamefont{C.}~\bibnamefont{{Hirata}}},
  \bibinfo{journal}{\prd} \textbf{\bibinfo{volume}{82}}, \bibinfo{eid}{083520}
  (\bibinfo{year}{2010}), \eprint{1005.2416}.

\bibitem[{\citenamefont{{Dalal} et~al.}(2010)\citenamefont{{Dalal}, {Pen}, and
  {Seljak}}}]{dalal/etal:2010}
\bibinfo{author}{\bibfnamefont{N.}~\bibnamefont{{Dalal}}},
  \bibinfo{author}{\bibfnamefont{U.-L.} \bibnamefont{{Pen}}}, \bibnamefont{and}
  \bibinfo{author}{\bibfnamefont{U.}~\bibnamefont{{Seljak}}},
  \bibinfo{journal}{\jcap} \textbf{\bibinfo{volume}{11}}, \bibinfo{eid}{007}
  (\bibinfo{year}{2010}), \eprint{1009.4704}.

\bibitem[{\citenamefont{{Yoo} et~al.}(2011)\citenamefont{{Yoo}, {Dalal}, and
  {Seljak}}}]{yoo/dalal/seljak}
\bibinfo{author}{\bibfnamefont{J.}~\bibnamefont{{Yoo}}},
  \bibinfo{author}{\bibfnamefont{N.}~\bibnamefont{{Dalal}}}, \bibnamefont{and}
  \bibinfo{author}{\bibfnamefont{U.}~\bibnamefont{{Seljak}}},
  \bibinfo{journal}{\jcap} \textbf{\bibinfo{volume}{7}}, \bibinfo{eid}{018}
  (\bibinfo{year}{2011}), \eprint{1105.3732}.

\bibitem[{\citenamefont{{Yoo} and {Seljak}}(2013)}]{yoo/seljak}
\bibinfo{author}{\bibfnamefont{J.}~\bibnamefont{{Yoo}}} \bibnamefont{and}
  \bibinfo{author}{\bibfnamefont{U.}~\bibnamefont{{Seljak}}},
  \bibinfo{journal}{\prd} \textbf{\bibinfo{volume}{88}}, \bibinfo{eid}{103520}
  (\bibinfo{year}{2013}), \eprint{1308.1401}.

\bibitem[{\citenamefont{{Blazek} et~al.}(2016)\citenamefont{{Blazek}, {McEwen},
  and {Hirata}}}]{blazek/etal:15}
\bibinfo{author}{\bibfnamefont{J.~A.} \bibnamefont{{Blazek}}},
  \bibinfo{author}{\bibfnamefont{J.~E.} \bibnamefont{{McEwen}}},
  \bibnamefont{and} \bibinfo{author}{\bibfnamefont{C.~M.}
  \bibnamefont{{Hirata}}}, \bibinfo{journal}{Physical Review Letters}
  \textbf{\bibinfo{volume}{116}}, \bibinfo{eid}{121303} (\bibinfo{year}{2016}),
  \eprint{1510.03554}.

\bibitem[{\citenamefont{{Slepian} and {Eisenstein}}(2015)}]{slepian/eisenstein}
\bibinfo{author}{\bibfnamefont{Z.}~\bibnamefont{{Slepian}}} \bibnamefont{and}
  \bibinfo{author}{\bibfnamefont{D.~J.} \bibnamefont{{Eisenstein}}},
  \bibinfo{journal}{\mnras} \textbf{\bibinfo{volume}{448}}, \bibinfo{pages}{9}
  (\bibinfo{year}{2015}), \eprint{1411.4052}.

\bibitem[{\citenamefont{{Mirbabayi} et~al.}(2015)\citenamefont{{Mirbabayi},
  {Schmidt}, and {Zaldarriaga}}}]{MSZ}
\bibinfo{author}{\bibfnamefont{M.}~\bibnamefont{{Mirbabayi}}},
  \bibinfo{author}{\bibfnamefont{F.}~\bibnamefont{{Schmidt}}},
  \bibnamefont{and}
  \bibinfo{author}{\bibfnamefont{M.}~\bibnamefont{{Zaldarriaga}}},
  \bibinfo{journal}{\jcap} \textbf{\bibinfo{volume}{7}}, \bibinfo{eid}{030}
  (\bibinfo{year}{2015}), \eprint{1412.5169}.

\bibitem[{\citenamefont{{Assassi} et~al.}(2014)\citenamefont{{Assassi},
  {Baumann}, {Green}, and {Zaldarriaga}}}]{assassi/etal}
\bibinfo{author}{\bibfnamefont{V.}~\bibnamefont{{Assassi}}},
  \bibinfo{author}{\bibfnamefont{D.}~\bibnamefont{{Baumann}}},
  \bibinfo{author}{\bibfnamefont{D.}~\bibnamefont{{Green}}}, \bibnamefont{and}
  \bibinfo{author}{\bibfnamefont{M.}~\bibnamefont{{Zaldarriaga}}},
  \bibinfo{journal}{\jcap} \textbf{\bibinfo{volume}{8}}, \bibinfo{eid}{056}
  (\bibinfo{year}{2014}), \eprint{1402.5916}.

\bibitem[{\citenamefont{{McDonald}}(2006)}]{mcdonald:2006}
\bibinfo{author}{\bibfnamefont{P.}~\bibnamefont{{McDonald}}},
  \bibinfo{journal}{\prd} \textbf{\bibinfo{volume}{74}}, \bibinfo{eid}{103512}
  (\bibinfo{year}{2006}), \eprint{arXiv:astro-ph/0609413}.

\bibitem[{\citenamefont{{Baumann} et~al.}(2012)\citenamefont{{Baumann},
  {Nicolis}, {Senatore}, and {Zaldarriaga}}}]{baumann}
\bibinfo{author}{\bibfnamefont{D.}~\bibnamefont{{Baumann}}},
  \bibinfo{author}{\bibfnamefont{A.}~\bibnamefont{{Nicolis}}},
  \bibinfo{author}{\bibfnamefont{L.}~\bibnamefont{{Senatore}}},
  \bibnamefont{and}
  \bibinfo{author}{\bibfnamefont{M.}~\bibnamefont{{Zaldarriaga}}},
  \bibinfo{journal}{\jcap} \textbf{\bibinfo{volume}{7}}, \bibinfo{eid}{051}
  (\bibinfo{year}{2012}), \eprint{1004.2488}.

\bibitem[{\citenamefont{Senatore}(2014)}]{senatore:2014}
\bibinfo{author}{\bibfnamefont{L.}~\bibnamefont{Senatore}}
  (\bibinfo{year}{2014}), \eprint{1406.7843}.

\bibitem[{\citenamefont{{Ahn}}(2016)}]{ahn}
\bibinfo{author}{\bibfnamefont{K.}~\bibnamefont{{Ahn}}},
  \bibinfo{journal}{ArXiv e-prints}  (\bibinfo{year}{2016}),
  \eprint{1603.09356}.

\bibitem[{\citenamefont{{Tseliakhovich}
  et~al.}(2011)\citenamefont{{Tseliakhovich}, {Barkana}, and
  {Hirata}}}]{tseliakhovich/barkana/hirata}
\bibinfo{author}{\bibfnamefont{D.}~\bibnamefont{{Tseliakhovich}}},
  \bibinfo{author}{\bibfnamefont{R.}~\bibnamefont{{Barkana}}},
  \bibnamefont{and} \bibinfo{author}{\bibfnamefont{C.~M.}
  \bibnamefont{{Hirata}}}, \bibinfo{journal}{\mnras}
  \textbf{\bibinfo{volume}{418}}, \bibinfo{pages}{906} (\bibinfo{year}{2011}),
  \eprint{1012.2574}.

\bibitem[{\citenamefont{{Visbal} et~al.}(2012)\citenamefont{{Visbal},
  {Barkana}, {Fialkov}, {Tseliakhovich}, and {Hirata}}}]{Visbal/etal:12}
\bibinfo{author}{\bibfnamefont{E.}~\bibnamefont{{Visbal}}},
  \bibinfo{author}{\bibfnamefont{R.}~\bibnamefont{{Barkana}}},
  \bibinfo{author}{\bibfnamefont{A.}~\bibnamefont{{Fialkov}}},
  \bibinfo{author}{\bibfnamefont{D.}~\bibnamefont{{Tseliakhovich}}},
  \bibnamefont{and} \bibinfo{author}{\bibfnamefont{C.~M.}
  \bibnamefont{{Hirata}}}, \bibinfo{journal}{\nat}
  \textbf{\bibinfo{volume}{487}}, \bibinfo{pages}{70} (\bibinfo{year}{2012}),
  \eprint{1201.1005}.

\bibitem[{\citenamefont{{Popa} et~al.}(2015)\citenamefont{{Popa}, {Naoz},
  {Marinacci}, and {Vogelsberger}}}]{popa/etal}
\bibinfo{author}{\bibfnamefont{C.}~\bibnamefont{{Popa}}},
  \bibinfo{author}{\bibfnamefont{S.}~\bibnamefont{{Naoz}}},
  \bibinfo{author}{\bibfnamefont{F.}~\bibnamefont{{Marinacci}}},
  \bibnamefont{and}
  \bibinfo{author}{\bibfnamefont{M.}~\bibnamefont{{Vogelsberger}}},
  \bibinfo{journal}{ArXiv e-prints}  (\bibinfo{year}{2015}),
  \eprint{1512.06862}.

\bibitem[{\citenamefont{{Lewis} et~al.}(2000)\citenamefont{{Lewis},
  {Challinor}, and {Lasenby}}}]{camb}
\bibinfo{author}{\bibfnamefont{A.}~\bibnamefont{{Lewis}}},
  \bibinfo{author}{\bibfnamefont{A.}~\bibnamefont{{Challinor}}},
  \bibnamefont{and}
  \bibinfo{author}{\bibfnamefont{A.}~\bibnamefont{{Lasenby}}},
  \bibinfo{journal}{\apj} \textbf{\bibinfo{volume}{538}}, \bibinfo{pages}{473}
  (\bibinfo{year}{2000}), \eprint{astro-ph/9911177}.

\bibitem[{\citenamefont{{Fry} and {Gaztanaga}}(1993)}]{fry/gaztanaga:1983}
\bibinfo{author}{\bibfnamefont{J.~N.} \bibnamefont{{Fry}}} \bibnamefont{and}
  \bibinfo{author}{\bibfnamefont{E.}~\bibnamefont{{Gaztanaga}}},
  \bibinfo{journal}{\apj} \textbf{\bibinfo{volume}{413}}, \bibinfo{pages}{447}
  (\bibinfo{year}{1993}), \eprint{arXiv:astro-ph/9302009}.

\bibitem[{\citenamefont{{Kaiser}}(1984)}]{kaiser:1984}
\bibinfo{author}{\bibfnamefont{N.}~\bibnamefont{{Kaiser}}},
  \bibinfo{journal}{\apjl} \textbf{\bibinfo{volume}{284}}, \bibinfo{pages}{L9}
  (\bibinfo{year}{1984}).

\bibitem[{\citenamefont{Bond et~al.}(1991)\citenamefont{Bond, Cole, Efstathiou,
  and Kaiser}}]{bond/etal:91}
\bibinfo{author}{\bibfnamefont{J.~R.} \bibnamefont{Bond}},
  \bibinfo{author}{\bibfnamefont{S.}~\bibnamefont{Cole}},
  \bibinfo{author}{\bibfnamefont{G.}~\bibnamefont{Efstathiou}},
  \bibnamefont{and} \bibinfo{author}{\bibfnamefont{N.}~\bibnamefont{Kaiser}},
  \bibinfo{journal}{The Astrophysical Journal} \textbf{\bibinfo{volume}{379}},
  \bibinfo{pages}{440} (\bibinfo{year}{1991}).

\bibitem[{\citenamefont{{Cole} and {Kaiser}}(1989)}]{cole/kaiser:1989}
\bibinfo{author}{\bibfnamefont{S.}~\bibnamefont{{Cole}}} \bibnamefont{and}
  \bibinfo{author}{\bibfnamefont{N.}~\bibnamefont{{Kaiser}}},
  \bibinfo{journal}{\mnras} \textbf{\bibinfo{volume}{237}},
  \bibinfo{pages}{1127} (\bibinfo{year}{1989}).

\bibitem[{\citenamefont{{Mo} and {White}}(1996)}]{mo/white:1996}
\bibinfo{author}{\bibfnamefont{H.~J.} \bibnamefont{{Mo}}} \bibnamefont{and}
  \bibinfo{author}{\bibfnamefont{S.~D.~M.} \bibnamefont{{White}}},
  \bibinfo{journal}{\mnras} \textbf{\bibinfo{volume}{282}},
  \bibinfo{pages}{347} (\bibinfo{year}{1996}).

\bibitem[{\citenamefont{{Asaba} et~al.}(2016)\citenamefont{{Asaba}, {Ichiki},
  and {Tashiro}}}]{asaba/ichiki/tashiro}
\bibinfo{author}{\bibfnamefont{S.}~\bibnamefont{{Asaba}}},
  \bibinfo{author}{\bibfnamefont{K.}~\bibnamefont{{Ichiki}}}, \bibnamefont{and}
  \bibinfo{author}{\bibfnamefont{H.}~\bibnamefont{{Tashiro}}},
  \bibinfo{journal}{\prd} \textbf{\bibinfo{volume}{93}}, \bibinfo{eid}{023518}
  (\bibinfo{year}{2016}), \eprint{1508.07719}.

\bibitem[{\citenamefont{{Ichiki} and {Takada}}(2012)}]{ichiki/takada}
\bibinfo{author}{\bibfnamefont{K.}~\bibnamefont{{Ichiki}}} \bibnamefont{and}
  \bibinfo{author}{\bibfnamefont{M.}~\bibnamefont{{Takada}}},
  \bibinfo{journal}{\prd} \textbf{\bibinfo{volume}{85}}, \bibinfo{eid}{063521}
  (\bibinfo{year}{2012}), \eprint{1108.4688}.

\bibitem[{\citenamefont{{Schmidt} et~al.}(2010)\citenamefont{{Schmidt}, {Hu},
  and {Lima}}}]{schmidt/hu/lima}
\bibinfo{author}{\bibfnamefont{F.}~\bibnamefont{{Schmidt}}},
  \bibinfo{author}{\bibfnamefont{W.}~\bibnamefont{{Hu}}}, \bibnamefont{and}
  \bibinfo{author}{\bibfnamefont{M.}~\bibnamefont{{Lima}}},
  \bibinfo{journal}{\prd} \textbf{\bibinfo{volume}{81}}, \bibinfo{eid}{063005}
  (\bibinfo{year}{2010}), \eprint{0911.5178}.

\bibitem[{\citenamefont{Angulo et~al.}(2015)\citenamefont{Angulo, Fasiello,
  Senatore, and Vlah}}]{angulo/etal:2015}
\bibinfo{author}{\bibfnamefont{R.}~\bibnamefont{Angulo}},
  \bibinfo{author}{\bibfnamefont{M.}~\bibnamefont{Fasiello}},
  \bibinfo{author}{\bibfnamefont{L.}~\bibnamefont{Senatore}}, \bibnamefont{and}
  \bibinfo{author}{\bibfnamefont{Z.}~\bibnamefont{Vlah}}
  (\bibinfo{year}{2015}), \eprint{1503.08826}.

\bibitem[{\citenamefont{{Assassi} et~al.}(2015)\citenamefont{{Assassi},
  {Baumann}, and {Schmidt}}}]{assassi/baumann/schmidt}
\bibinfo{author}{\bibfnamefont{V.}~\bibnamefont{{Assassi}}},
  \bibinfo{author}{\bibfnamefont{D.}~\bibnamefont{{Baumann}}},
  \bibnamefont{and}
  \bibinfo{author}{\bibfnamefont{F.}~\bibnamefont{{Schmidt}}},
  \bibinfo{journal}{\jcap} \textbf{\bibinfo{volume}{12}}, \bibinfo{eid}{043}
  (\bibinfo{year}{2015}), \eprint{1510.03723}.

\bibitem[{\citenamefont{Tellarini et~al.}(2015)\citenamefont{Tellarini, Ross,
  Tasinato, and Wands}}]{tellarini/etal}
\bibinfo{author}{\bibfnamefont{M.}~\bibnamefont{Tellarini}},
  \bibinfo{author}{\bibfnamefont{A.}~\bibnamefont{Ross}},
  \bibinfo{author}{\bibfnamefont{G.}~\bibnamefont{Tasinato}}, \bibnamefont{and}
  \bibinfo{author}{\bibfnamefont{D.}~\bibnamefont{Wands}},
  \bibinfo{journal}{JCAP} \textbf{\bibinfo{volume}{1507}}, \bibinfo{pages}{004}
  (\bibinfo{year}{2015}), \eprint{1504.00324}.

\bibitem[{\citenamefont{{McDonald} and {Roy}}(2009)}]{mcdonald/roy:2009}
\bibinfo{author}{\bibfnamefont{P.}~\bibnamefont{{McDonald}}} \bibnamefont{and}
  \bibinfo{author}{\bibfnamefont{A.}~\bibnamefont{{Roy}}},
  \bibinfo{journal}{\jcap} \textbf{\bibinfo{volume}{8}}, \bibinfo{pages}{20}
  (\bibinfo{year}{2009}), \eprint{0902.0991}.

\bibitem[{\citenamefont{{Hamilton}}(1998)}]{hamilton:1998}
\bibinfo{author}{\bibfnamefont{A.~J.~S.} \bibnamefont{{Hamilton}}}, in
  \emph{\bibinfo{booktitle}{The Evolving Universe}}, edited by
  \bibinfo{editor}{\bibfnamefont{D.}~\bibnamefont{{Hamilton}}}
  (\bibinfo{year}{1998}), vol. \bibinfo{volume}{231} of
  \emph{\bibinfo{series}{Astrophysics and Space Science Library}}, p.
  \bibinfo{pages}{185}, \eprint{astro-ph/9708102}.

\bibitem[{\citenamefont{Kaiser}(1987)}]{kaiser:1987}
\bibinfo{author}{\bibfnamefont{N.}~\bibnamefont{Kaiser}},
  \bibinfo{journal}{Monthly Notices of the Royal Astronomical Society (ISSN
  0035-8711)} \textbf{\bibinfo{volume}{227}}, \bibinfo{pages}{1}
  (\bibinfo{year}{1987}).

\bibitem[{\citenamefont{{Slepian} et~al.}(2016)\citenamefont{{Slepian},
  {Eisenstein}, {Blazek}, {Brownstein}, {Chuang}, {Gil-Mar{\'{\i}}n}, {Ho},
  {Kitaura}, {McEwen}, {Percival} et~al.}}]{slepian/etal:2016}
\bibinfo{author}{\bibfnamefont{Z.}~\bibnamefont{{Slepian}}},
  \bibinfo{author}{\bibfnamefont{D.~J.} \bibnamefont{{Eisenstein}}},
  \bibinfo{author}{\bibfnamefont{J.~A.} \bibnamefont{{Blazek}}},
  \bibinfo{author}{\bibfnamefont{J.~R.} \bibnamefont{{Brownstein}}},
  \bibinfo{author}{\bibfnamefont{C.-H.} \bibnamefont{{Chuang}}},
  \bibinfo{author}{\bibfnamefont{H.}~\bibnamefont{{Gil-Mar{\'{\i}}n}}},
  \bibinfo{author}{\bibfnamefont{S.}~\bibnamefont{{Ho}}},
  \bibinfo{author}{\bibfnamefont{F.-S.} \bibnamefont{{Kitaura}}},
  \bibinfo{author}{\bibfnamefont{J.~E.} \bibnamefont{{McEwen}}},
  \bibinfo{author}{\bibfnamefont{W.~J.} \bibnamefont{{Percival}}},
  \bibnamefont{et~al.}, \bibinfo{journal}{ArXiv e-prints}
  (\bibinfo{year}{2016}), \eprint{1607.06098}.

\bibitem[{\citenamefont{Schmidt et~al.}(2009)\citenamefont{Schmidt, Lima,
  Oyaizu, and Hu}}]{HPMhalopaper}
\bibinfo{author}{\bibfnamefont{F.}~\bibnamefont{Schmidt}},
  \bibinfo{author}{\bibfnamefont{M.~V.} \bibnamefont{Lima}},
  \bibinfo{author}{\bibfnamefont{H.}~\bibnamefont{Oyaizu}}, \bibnamefont{and}
  \bibinfo{author}{\bibfnamefont{W.}~\bibnamefont{Hu}}, \bibinfo{journal}{Phys.
  Rev.} \textbf{\bibinfo{volume}{D79}}, \bibinfo{pages}{083518}
  (\bibinfo{year}{2009}), \eprint{0812.0545}.

\end{thebibliography}

\end{document}